\documentclass[a4paper,11pt]{article}
\usepackage{amsfonts}
\usepackage{cite}

\usepackage{graphicx}
\graphicspath{{Figures/}}
\usepackage{amsmath}
\usepackage{amssymb}
\usepackage{latexsym}
\usepackage[colorlinks]{hyperref}
\usepackage{color}
\usepackage{float}
\usepackage{cite}
\usepackage{makeidx}
\usepackage{colortbl}
\usepackage{csquotes}
\usepackage[squaren]{SIunits}

\usepackage[textfont={footnotesize,sf},labelfont={color=blue,bf,sf},labelsep=endash]{caption}
\usepackage[position=top,labelfont={color=blue,bf,sf}]{subfig}
\usepackage{bm}

\usepackage[title]{appendix}

\newcommand{\bra}{\begin{array}}
    \newcommand{\era}{\end{array}}
\newcommand{\beq}{\begin{equation}}
    \newcommand{\eeq}{\end{equation}}
\newcommand{\bqr}{\begin{eqnarray}}
    \newcommand{\eqr}{\end{eqnarray}}

\def\BC{\bb C}
\def\_\BC{\bbi C}

\def \lp {\left(}
\def \gp {\right)}

\def\no2 {{\textstyle{n\over 2}}}

\newcommand{\lb}{\label}

\begin{document}
\begin{titlepage}
\setcounter{page}{1}
\renewcommand{\thefootnote}{\fnsymbol{footnote}}

\begin{flushright}
\end{flushright}

\vspace{5mm}
\begin{center}

{\Large \bf { Tunneling of Electrons in Graphene
via Double Triangular Barrier in External
Fields}}

\vspace{5mm}
{\bf Miloud Mekkaoui}$^{a}$,
 {\bf Ahmed Jellal\footnote{\sf ajellal@ictp.it --
a.jellal@ucd.ac.ma}}$^{a,b}$ and {\bf Hocine Bahlouli}$^{c}$

\vspace{5mm}

{$^{a}$\em Laboratory of Theoretical Physics,  
Faculty of Sciences, Choua\"ib Doukkali University},\\
{\em PO Box 20, 24000 El Jadida, Morocco}

{$^{b}$\em Canadian Quantum  Research Center,
	204-3002 32 Ave Vernon, \\ BC V1T 2L7,  Canada}


{$^c$\em Physics Department,  King Fahd University
of Petroleum $\&$ Minerals,\\
Dhahran 31261, Saudi Arabia}

\vspace{3cm}

\begin{abstract}
We study the transmission probability of Dirac fermions in
graphene scattered by a triangular double barrier potential in the
presence of an external magnetic field. Our system made of two
triangular potential barrier regions separated by a well region
characterized by an energy gap. Solving our Dirac-like
equation and matching the solutions at the boundaries allowed us to express
our transmission and reflection coefficients in terms of transfer
matrix. We show in particular that the transmission
exhibits oscillation resonances that are manifestations of the
Klein tunneling effect. The triangular barrier electrostatic field was found to
play a key role in controlling the peak of tunneling resistance. However, it
only slightly modifies the resonances at oblique incidence and leaves Klein paradox unaffected at normal incidence.

\end{abstract}
\end{center}

\vspace{3cm}


\end{titlepage}


\section{Introduction}

Graphene~\cite{Novoselov} remains among the most fascinating and
attractive subjects in contemporary condensed matter physics. This is
because of its exotic physical and transport properties and the apparent
similarity of its mathematical model to the one describing massless
relativistic fermions in two dimensions. As a consequence of this
relativistic-like behavior, particles could tunnel through very high
barriers in contrast to the conventional tunneling of
non-relativistic particles, an effect known in relativistic field
theory as Klein tunneling. This tunneling effect has already been
observed experimentally~\cite{Stander} in graphene systems. However, there
are various ways for creating barrier structures in
graphene~\cite{Katsnelsonn, Sevincli}. For instance, it can be realized
by applying a gate voltage, cutting the graphene sheet into a finite
width to create a nanoribbons, using doping or through the
creation of a magnetic barrier. In the case of graphene, results
of the transmission coefficient and the tunneling conductance were
already reported for the electrostatic barriers~\cite{
Masir, Dell'Anna, Mukhopadhyay}, magnetic barriers~\cite{Dell'Anna, Choubabi, Mekkaoui},
potential barrier~\cite{Jellal}, linear~\cite{HBahlouli}
and triangular \cite{Elmouhafid2013} barriers.

One of the present author \cite{Jellal} 
 theoretically studied the electronic transport properties of Dirac fermions
through one and double triangular barriers in graphene nanoribbon. 
The transmission, conductance and Fano factor
are obtained to be various parameters dependent such as well width, barrier height
and barrier width. Therefore, different discussions are given and comparison with
the previous significant works is done. In particular, it is shown that at Dirac point
the Dirac fermions always own a minimum conductance associated with a maximum
Fano factor and change their behaviors in an oscillatory way (irregularly periodical
tunneling peaks) when the potential of applied voltage is increased.

In our present work we study the transmission probability of Dirac fermions in graphene scattered by a triangular double barrier potential in the presence of a uniform magnetic fields $B$ confined to the well region between the two barriers. We emphasis that  $B$-field discussed in our manuscript is applied externally and is perpendicular to the graphene sheet. It can be created for instance by depositing a type-I superconducting film on top of the
system except for a strip $|x|<d_1$ which is isolated in order to
apply a perpendicular magnetic field in its domain. This patterning technique of
creating the desired magnetic field profile was proposed in
\cite{Matulis}. One of the interesting features of such a
magnetic field profile is that it can bind
electrons, contrary to the usual potential step profile. Such a step
magnetic field will indeed result in electron states that are
bound to the step $B$-field and that move in one direction along the
step. Thus there is a current along the $y$-direction but it is a
very small effect and is not relevant for our problem (those
electrons have $k_{x} = 0$). Indeed, we consider free electron
states that have in general non zero $k_x$, because otherwise they
will not tunnel. A recent work studied double barriers with
magnetic field in graphene without mass term \cite{Ramezani}.



The paper is organized as follows. In section 2, we formulate our
model by setting the Hamiltonian system describing particles
scattered by a triangular double barrier whose well potential
zone is subject to a magnetic field in the presence of a mass term.
In section 3, we consider the case of static double barriers and derive the
energy spectrum to finally determine the transmission and
reflection probabilities. Their behaviors are numerically
investigated and 
in particular resonances were seen
in different regions while Klein tunneling was observed to persist at normal incidence.
In section 4, we study the same system but
this time taking into account the presence of a uniform magnetic field
confined to the barrier region. 
Using boundary conditions, we  split the energy into three domains
and then calculate the transmission probability in each case.
In each situation, we discuss the transmission resonances that
characterize each region and stress the importance of our
results. We conclude our work in the final section.

\section{ System setting}

\begin{figure}[ht]
	\centering
	\includegraphics[width=10cm, height=4cm ]{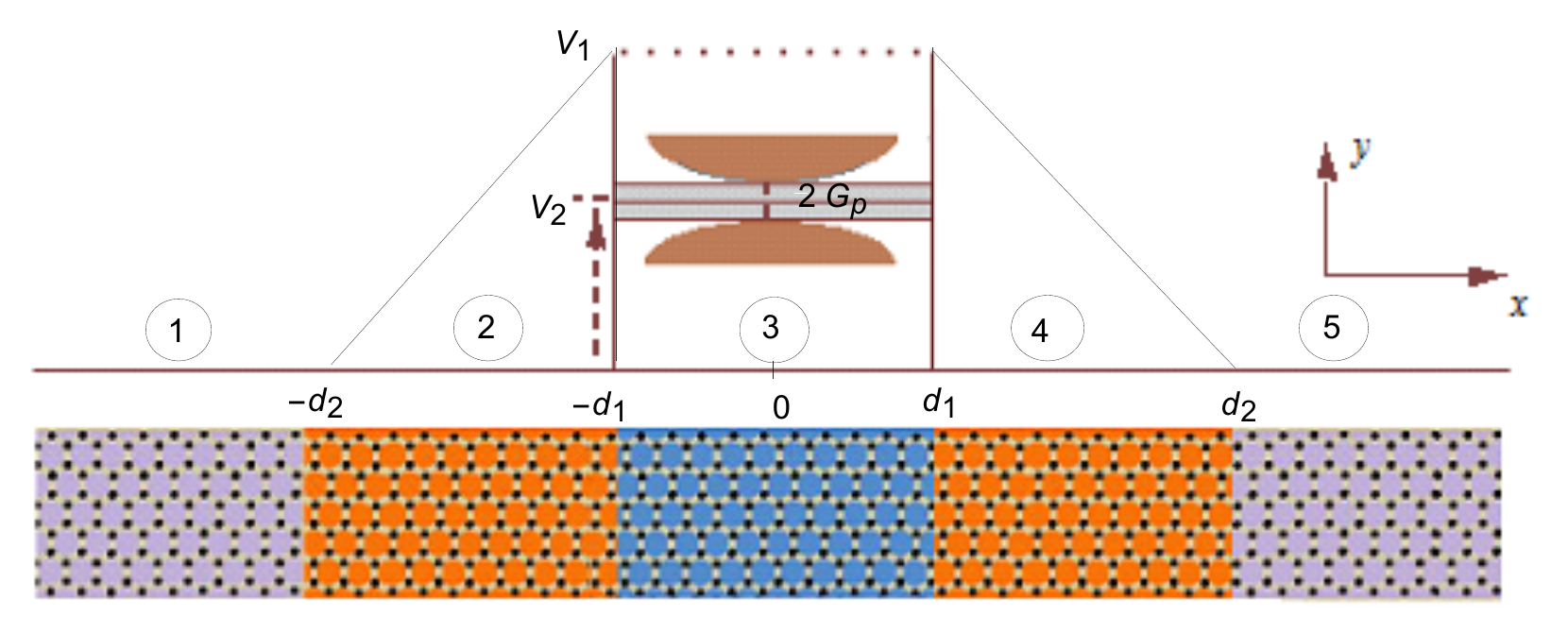}\\
	\caption{\sf Schematic diagram for the monolayer graphene double barrier.}\label{db.1}
\end{figure}

Our system
is a flat sheet of graphene subject to  the barrier $V(x)=V_{\sf j}$
along the $x$-direction while particles are free in the
$y$-direction and a mass term is applied in region 3 as graphically 
represented  in
Figure \ref{db.1}. Formally, we have  regions ${\sf{j = 1}}, {\sf{\cdots}}, {{\sf5}}$
characterized by
 the electrostatic potential 
\begin{equation}\lb{popro}
	V(x)=
	\left\{%
	\begin{array}{ll}
		\Lambda ( d_2 + \gamma x ) , & \hbox{$d_{1}\leq |x|\leq d_{2}$} \\
		V_{2}, & \hbox{$ |x|\leq d_{1}$} \\
		0, & \hbox{otherwise} \\
	\end{array}%
	\right.
\end{equation}
where $\gamma=1$ for $x\in [-d_2, -d_1]$, $\gamma=-1$ for $x\in
[d_1, d_2]$ and the parameter $ \Lambda =
\frac{V_1}{d_2-d_1}$ gives the slope of triangular potentials
which represents the strength of the applied electrostatic filed.

The Hamiltonian system can be written as 
\begin{equation}\lb{Ham1}
H=v_{F}
{\boldsymbol{\sigma}}\cdot\left(\textbf{p}+\frac{e}{c}\textbf{A}\right)+
V(x){\mathbb
I}_{2}+G_p\Theta\left(d_{1}^{2}-x^{2}\right)\sigma_{z}
\end{equation}
with the Fermi
velocity
 {${v_{F}\approx 10^6 m/s}$,
 	the Pauli matrices $\sigma_{i}$
 	%
in pseudospin space,
the momentum operator
$\textbf{p}=(p_{x},
p_{y})$,  the $2 \times 2$ unit matrix ${\mathbb I}_{2}$
and $\Theta$ is the Heaviside step function. The  vector potential
associated to perpendicular constant magnetic field $B$ will be chosen 
the Landau gauge $\textbf{A} =
(0,Bx)$.
The energy gap $G_p = m v_{F}^2$ is originating from the sublattice symmetry breaking, which also can be
seen as 
$G_p = G_{p, so}$ originating from spin-orbit interaction.


\section{Static double barrier}

\subsection{Energy spectrum: electrostatic double barrier}


We consider a Hamiltonian describing Dirac fermions in graphene scattered by an electrostatic double barrier potential in the absence of magnetic field, $\textbf{A} = 0$. In this case \eqref{Ham1} reduces to
\begin{equation} \lb{eqh1}
H_{s}=v_{F} {\boldsymbol{\sigma}}\cdot\textbf{p}+V (x){\mathbb
I}_{2}+G_p\Theta\left(d_{1}^{2}-x^{2}\right)\sigma_{z}
\end{equation}
where ${\sf{j}}$ labels the five regions indicated schematically in
Figure \ref{db.1}  showing the space configuration of the
potential profile.  Due to inversion and time reversal symmetries we therefore need to study our system only near the \textbf{K} point. The time-independent Dirac equation for the spinor
$\Phi(x,y)=\left(\varphi^{+},\varphi^{-}\right)^{T}$ at energy
$E=v_{F}\epsilon$ then reads, in unit system $\hbar = m = c= 1$, as
\begin{equation} \lb{eqh2}
\left[{\boldsymbol{\sigma}}\cdot\textbf{p}+{v_j}{\mathbb
I}_{2}+\mu\Theta\left(d_{1}^{2}-x^{2}\right)\sigma_{z}\right]\Phi(x,y)=\epsilon
\Phi(x,y)
\end{equation}
where {$V_{\sf j}=v_{F}v_{\sf j}$ and $G_{p}=v_{F}\mu$}. Our
system is supposed to have finite width $W$ with infinite mass
boundary conditions on the wavefunction at the boundaries $y = 0$
and $y = W$ along the $y$-direction \cite{Tworzydlo, Berry}. These
boundary conditions result in a quantization of the transverse
momentum along the $y$-direction as
\begin{equation}
k_{y}=\frac{\pi}{W}\left(n+\frac{1}{2}\right),\qquad n=0,1,2 \cdots.
\end{equation}

One can therefore assume a spinor solution of the following form
$\Phi_{\sf j}=\left(\varphi_{\sf j}^{+}(x),\varphi_{\sf
j}^{-}(x)\right)^{\dagger}e^{ik_{y}y}$ and the subscript ${\sf
j}= 1, 2, 3, 4, 5$ refers to the space region while the
superscripts indicate the two spinor components. Solving the
eigenvalue equation to obtain the upper and lower components of
the eignespinor in the incident and reflection region {\sf 1} ($x
< - d_{2}$) gives
\begin{eqnarray}\label{eq3}
     \Phi_{\sf 1}=  \left(
            \begin{array}{c}
              {1} \\
              {z_{1}} \\
            \end{array}
          \right) e^{i(k_{1}x+k_{y}y)} + r_{s,n}\left(
            \begin{array}{c}
              {1} \\
              {-z_{1}^{-1}} \\
            \end{array}
          \right) e^{i(-k_{1}x+k_{y}y)}, \qquad 
 z_{1} =s_{1}\frac{k_{1}+ik_{y}}{\sqrt{k_{1}^{2}+k_{y}^{2}}}
\end{eqnarray}
associated to the  dispersion relation $ \epsilon=s_1\sqrt{k_1^2 +k_y^2} $
with 
$s_{\sf j}={\mbox{sign}}{\left(E\right)}$.

In regions {\sf 2} and {\sf 4} ($d_{1}<|x|<d_{2}$),  the general
solution can be expressed in terms of the parabolic cylinder
function \cite{Abramowitz, Gonzalez, HBahlouli} as
\begin{equation}\lb{hiii1}
 \chi_{\gamma}^{+}=c_{n1}
 D_{\nu_n-1}\left(Q_{\gamma}\right)+c_{n2}
 D_{-\nu_n}\left(-Q^{*}_{\gamma}\right)
\end{equation}
where {$c_{n1}$ and $c_{n2}$ are constants,
$\nu_n=\frac{ik_{y}^{2}}{2\varrho}$ and $
Q_{\gamma}(x)=\sqrt{\frac{2}{\varrho}}e^{i\pi/4}\left(\gamma
\varrho x+\epsilon_{0}\right)$, with
$\epsilon_{0}=\epsilon-v_{1}$, $\Lambda=v_{F}\varrho$,
$V_{1}=v_{F}v_{1}$}. The lower spinor component is given by
\begin{eqnarray}\lb{hiii2}
\chi_{\gamma}^{-}=-\frac{c_{n2}}{k_{y}}\left[
2(\epsilon_{0}+\gamma \varrho x)
 D_{-\nu_n}\left(-Q^{*}_{\gamma}\right)
+
 \sqrt{2\varrho}e^{i\pi/4}D_{-\nu_n+1}\left(-Q^{*}_{\gamma}\right)\right]
 -\frac{c_{n1}}{k_{y}}\sqrt{2\varrho}e^{-i\pi/4}
 D_{\nu_n-1}\left(Q_{\gamma}\right).
\end{eqnarray}
The components of the spinor solution of the Dirac equation
\eqref{eqh1} in regions {\sf 2} and {\sf 4} can be obtained from
\eqref{hiii1} and \eqref{hiii2} with
$\varphi_{\gamma}^{+}(x)=\chi_{\gamma}^{+}+i\chi_{\gamma}^{-}$ and
$\varphi_{\gamma}^{-}(x)=\chi_{\gamma}^{+}-i\chi_{\gamma}^{-}$. Then, in
regions {\sf 2} and {\sf 4}
we have the eigenspinors
\begin{eqnarray}
 \Phi_{\sf j } &=& a_{\sf{j}-1}\left(%
\begin{array}{c}
 \eta^{+}_{\gamma}(x) \\
  \eta^{-}_{\gamma}(x) \\
\end{array}%
\right)e^{ik_{y}y}+a_{\sf j}\left(%
\begin{array}{c}
 \xi^{+}_{\gamma}(x) \\
 \xi^{-}_{\gamma}(x)\\
\end{array}%
\right)e^{ik_{y}y}
\end{eqnarray}
where $\gamma=\pm 1$ and we have set
\begin{eqnarray}
&& \eta^{\pm}_{\gamma}(x)=
 D_{\nu_{n}-1}\left(Q_{\gamma}\right)\mp
 \frac{1}{k_{y}}\sqrt{2\varrho}e^{i\pi/4}D_{\nu_{n}}\left(Q_{\gamma}\right)\\
&& \xi^{\pm}_{\gamma}(x)=
 \pm\frac{1}{k_{y}}\sqrt{2\varrho}e^{-i\pi/4}D_{-\nu_{n}+1}\left(-Q_{\gamma}^{*}\right)
  \pm
 \frac{1}{k_{y}}\left(-2i\epsilon_{0}\pm
 k_{y}-\gamma2i \varrho x\right)D_{-\nu_{n}}\left(-Q_{\gamma}^{*}\right).
\end{eqnarray}

In region 3, we use \eqref{eqh1}  to obtain
the eigenspinor
\begin{eqnarray} \label{eq 7}
  \Phi_{\sf 3}= b_1 \left(
            \begin{array}{c}
              {\alpha} \\
              {\beta z_{3}} \\
            \end{array}
          \right) e^{i(k_{3}x+k_{y}y)} +b_2 \left(
            \begin{array}{c}
              {\alpha} \\
              {-\beta z_{3}^{-1}} \\
            \end{array}
          \right) e^{i(-k_{3}x+k_{y}y)}
        \end{eqnarray}
    with the parameters 
    \begin{equation}\label{1313}
    	z_{3}
    	=s_{3}\frac{k_{3}+ik_{y}}{\sqrt{k_{3}^{2}+k_{y}^{2}}}, 
    	\qquad 
       {\alpha=\lp{1+\frac{\mu}{ \epsilon-v_{2}}}\gp}^{1/2}, \qquad
       {\beta=\lp{1-\frac{\mu}{ \epsilon-v_{2}}}\gp}^{1/2}
 \end{equation}
and the wave vector $ k_{3}= \sqrt{(\epsilon-v_{2})^{2}-\mu^{2}-{k_{y}}^{2} }$,
 $s_{3}=\mbox{sign}(\epsilon-v_{2})$.
The eigenspinor in region {\sf 5} can easily be obtained
\begin{equation}\label{eq6}
 \Phi_{\sf 5}= t_{s,n} \left(
            \begin{array}{c}
              {1} \\
              {z_{1}} \\
            \end{array}
          \right) e^{i(k_{1}x+k_{y}y)}
\end{equation}
in similar way to region {\sf 1}. These results will be embedded to deal with 
tunneling properties of our system. 

\subsection{ Transmission and reflection: electrostatic double barrier }

We determine the transmission and reflection probabilities 
for an electrostatic double barrier alone
 by requiring the continuity of the eigenspinor  at 
junction interfaces. This yields 
\bqr
\label{eq11} \Phi_{\sf 1}(-d_2)= \Phi_{\sf
2}(-d_2), \qquad \Phi_{\sf 2}(-d_1)= \Phi_{\sf 3}(-d_1), \qquad
\Phi_{\sf 3}(d_1)= \Phi_{\sf 4}(d_1), \qquad
\Phi_{\sf 4}(d_2)= \Phi_{\sf 5}(d_2).
\eqr
We prefer to express these relationships in terms of $2\times 2$
transfer matrices between different regions, then we write
\beq
\left(%
\begin{array}{c}
  a_{\sf j} \\
  b_{\sf j} \\
\end{array}%
\right)=M_{{\sf j}, {\sf j}+1}\left(%
\begin{array}{c}
  a_{{\sf j}+1} \\
  b_{{\sf j}+1} \\
\end{array}%
\right)
\eeq
where
$M_{{\sf j}, {\sf j}+1}$ are the transfer matrices that couple the
wavefunction in the $j$-th region to the wavefunction in the
$({\sf j} + 1)$-th region. Finally, we obtain the full transfer matrix $M$ over the
whole double barrier which can be written, in an obvious notation,
as follows
\begin{equation}\label{systm1}
\left(%
\begin{array}{c}
  1 \\
  r_{s,n} \\
\end{array}%
\right)=\prod_{{\sf j}=1}^{4}M_{{\sf j}, {\sf j}+1}\left(%
\begin{array}{c}
  t_{s,n} \\
  0 \\
\end{array}%
\right)=M\left(%
\begin{array}{c}
  t_{s,n} \\
  0 \\
\end{array}%
\right)
\end{equation}
where 
$M=M_{12}\cdot M_{2
3}\cdot M_{34}\cdot M_{45}$ is given by
\begin{equation}
 M=\left(%
\begin{array}{cc}
  m_{11} & m_{12} \\
  m_{21} & m_{22} \\
\end{array}%
\right)
\\
\end{equation}
and the remaining matrices read as
\begin{eqnarray}
&& M_{12}=\left(%
\begin{array}{cc}
   e^{-\textbf{\emph{i}}k_{1} d_{2}} &e^{\textbf{\emph{i}}k_{1} d_{2}} \\
  z_{1}e^{-\textbf{\emph{i}}k_{1} d_{2}} & -z^{\ast}_{1} e^{\textbf{\emph{i}}k_{1} d_{2}} \\
\end{array}%
\right)^{-1}\left(%
\begin{array}{cc}
\eta_{1}^{+}(-d_2) &  \xi_{1}^{+}(-d_2)\\
 \eta_{1}^{-}(-d_2) & \xi_{1}^{-} (-d_2)\\
\end{array}%
\right)
\\
&& M_{23}=\left(%
\begin{array}{cc}
 \eta_{1}^{+}(-d_1) &  \xi_{1}^{+}(-d_1)\\
 \eta_{1}^{-}(-d_1) & \xi_{1}^{-} (-d_1)\\
\end{array}%
\right)^{-1}\left(%
\begin{array}{cc}
\alpha e^{-\textbf{\emph{i}}k_{3} d_{1}} &\alpha e^{\textbf{\emph{i}}k_{3} d_{1}} \\
  \beta z_{3}e^{-\textbf{\emph{i}}k_{3} d_{1}} & -\beta z^{\ast}_{3} e^{\textbf{\emph{i}}k_{3} d_{1}} \\
\end{array}%
\right)
\end{eqnarray}
\begin{eqnarray}
&& M_{34}=\left(%
\begin{array}{cc}
 \alpha e^{\textbf{\emph{i}}k_{3} d_{1}} &\alpha e^{-\textbf{\emph{i}}k_{3} d_{1}} \\
  \beta z_{3}e^{\textbf{\emph{i}}k_{3} d_{1}} & -\beta z^{\ast}_{3} e^{-\textbf{\emph{i}}k_{3} d_{1}} \\
\end{array}%
\right)^{-1}\left(%
\begin{array}{cc}
 \eta_{-1}^{+}(d_1) &  \xi_{-1}^{+}(d_1)\\
 \eta_{-1}^{-}(d_1) & \xi_{-1}^{-} (d_1)\\
\end{array}%
\right)
\\
&& M_{45}=\left(%
\begin{array}{cc}
 \eta_{-1}^{+}(d_2) &  \xi_{-1}^{+}(d_2)\\
 \eta_{-1}^{-}(d_2) & \xi_{-1}^{-} (d_2)\\
\end{array}%
\right)^{-1}\left(%
\begin{array}{cc}
  e^{\textbf{\emph{i}}k_{1} d_{2}} & e^{-\textbf{\emph{i}}k_{1} d_{2}} \\
  z_{1} e^{\textbf{\emph{i}}k_{1} d_{2}}  & -z_{1}^{\ast} e^{-\textbf{\emph{i}}k_{1} d_{2}}  \\
\end{array}%
\right).
\end{eqnarray}
Consequently, the transmission and  reflection amplitudes take the forms
\begin{equation}\label{eq 63}
 t_{s,n}=\frac{1}{m_{11}}, \qquad  r_{s,n}=\frac{m_{21}}{m_{11}}.
\end{equation}
Some symmetry relationship between the parabolic cylindric functions are worth mentioning, they are given by
\begin{equation}
\eta_{-1}^{\pm}(d_1)=\eta_{1}^{\pm}(-d_1),\qquad
\eta_{-1}^{\pm}(d_2)=\eta_{1}^{\pm}(-d_2)
\end{equation}
\begin{equation}
 \xi_{-1}^{\pm}(d_1)=\xi_{1}^{\pm}(-d_1),\qquad
 \xi_{-1}^{\pm}(d_2)=\xi_{1}^{\pm}(-d_2).
\end{equation}

We should point out at this stage that we were unfortunately
forced to adopt a somehow cumbersome notation for our wavefunction
parameters in different potential regions due to the relatively
large number of necessary subscripts and superscripts. Before
matching the eigenspinors at the boundaries, let us define the
following shorthand notation
\begin{equation}
\eta_{1}^{\pm}(-d_1)=\eta_{11}^{\pm},\qquad
 \eta_{1}^{\pm}(-d_2)=\eta_{12}^{\pm}
\end{equation}
\begin{equation}
 \xi_{1}^{\pm}(-d_1)=\xi_{11}^{\pm},\qquad
 \xi_{1}^{\pm}(-d_2)=\xi_{12}^{\pm}.
\end{equation}
After
some lengthy algebra, one can solve the linear system given in
\eqref{systm1} to obtain the transmission and reflection
amplitudes in closed forms. As for the former one, we find
\begin{equation}
t_{s,n}=\frac{\alpha\beta e^{2i(k_{1}d_{2}+k_{3}d_{1})}
\left(1+z_{1}^{2}\right)\left(1+z_{3}^{2}\right)}{z_{3}\left(e^{4ik_{3}d_{1}}-1\right)\left(
\alpha^{2}\mathcal{Y}_{2}+\beta^{2}\mathcal{Y}_{1}\right)+\alpha\beta
\mathcal{Y}_{3}}\left(\xi_{11}^{+}\eta_{11}^{-}-\xi_{11}^{-}\eta_{11}^{+}\right)
\left(\xi_{12}^{-}\eta_{12}^{+}-\xi_{12}^{+}\eta_{12}^{-}\right)
 \end{equation}
where we have defined the following quantities
\begin{eqnarray}
 &&\mathcal{Y}_{1}=\left(\xi_{12}^{-}\eta_{11}^{+}-\xi_{11}^{+}\eta_{12}^{-}-
 \xi_{12}^{+}\eta_{11}^{+}z_{1}+\xi_{11}^{+}\eta_{12}^{+}z_{1}\right)\left( \xi_{11}^{+}\eta_{12}^{+}+
 \xi_{11}^{+}\eta_{12}^{-}z_{1}-\eta_{11}^{+}(\xi_{12}^{+}+\xi_{12}^{-}z_{1}\right)\\
 &&
 \mathcal{Y}_{2}=\left(\xi_{11}^{-}\eta_{12}^{+}-\xi_{11}^{-}\eta_{12}^{-}z_{1}-\eta_{11}^{-}(
 \xi_{12}^{+}+\xi_{12}^{+}z_{1}\right)
 \left( -\xi_{12}^{-}\eta_{11}^{-}+
 \xi_{12}^{+}\eta_{11}^{-}z_{1}-\xi_{11}^{-}(\eta_{12}^{-}+\eta_{12}^{+}z_{1}\right)\\
 &&
  \mathcal{Y}_{3}=\Gamma_{0}\left(1+z_{1}^{2}z_{3}^{2}\right)+\Gamma_{1}z_{1}\left(1-z_{3}\right)+\Gamma_{2}\left(z_{1}^{2}+z_{3}^{2}\right)
  +e^{4id_{1}k_{3}}\left(\Gamma_{3}+\Gamma_{4}\right)
\end{eqnarray}
together with
 \begin{eqnarray}
 \Gamma_{0}&=&-\xi_{12}^{+}\xi_{12}^{-}\eta_{11}^{+}\eta_{11}^{-}
 +\xi_{11}^{+}\xi_{12}^{-}\eta_{11}^{-}\eta_{12}^{+}+
 \xi_{11}^{-}\xi_{12}^{+}\eta_{11}^{+}\eta_{12}^{-}-
 \xi_{11}^{+}\xi_{11}^{-}\eta_{12}^{+}\eta_{12}^{-}\\
 \Gamma_{1}&=&\left(\xi_{12}^{+}\right)^{2}\eta_{11}^{+}\eta_{11}^{-}
 -\left(\xi_{12}^{-}\right)^{2}\eta_{11}^{+}\eta_{11}^{-}-
 \xi_{11}^{-}\xi_{12}^{+}\eta_{11}^{+}\eta_{12}^{+}-
 \xi_{11}^{+}\xi_{12}^{+}\eta_{11}^{-}\eta_{12}^{+}\\\nonumber
 &&
 +\xi_{11}^{+}\xi_{11}^{-}\left(\eta_{12}^{+}\right)^{2}
 -\xi_{11}^{+}\xi_{11}^{-}\left(\eta_{12}^{-}\right)^{2}+
 \xi_{11}^{-}\xi_{12}^{-}\eta_{11}^{+}\eta_{12}^{-}+
 \xi_{11}^{+}\xi_{12}^{-}\eta_{11}^{-}\eta_{12}^{-}\\
 \Gamma_{2}&=&\xi_{12}^{+}\xi_{12}^{-}\eta_{11}^{+}\eta_{11}^{-}
 -\xi_{11}^{-}\xi_{12}^{-}\eta_{11}^{+}\eta_{12}^{+}-
 \xi_{11}^{+}\xi_{12}^{+}\eta_{11}^{-}\eta_{12}^{-}+
 \xi_{11}^{+}\xi_{11}^{-}\eta_{12}^{+}\eta_{12}^{-}\\
 \Gamma_{3}&=&\left(\xi_{12}^{+}\right)^{2}\eta_{11}^{+}\eta_{11}^{-}\left(z_{3}^{2}-1\right)
 -\xi_{11}^{-}\xi_{12}^{-}\eta_{11}^{+}\left[\eta_{12}^{+}\left(1+z_{1}^{2}z_{3}^{2}\right)-\eta_{12}^{-}z_{1}\left(z_{3}^{2}-1\right)\right]\\\nonumber
 &&
 +\xi_{11}^{-}\xi_{11}^{+}\left[\left(\eta_{12}^{+}\right)^{2}z_{1}
 -\left(\eta_{12}^{-}\right)^{2}z_{1}+\eta_{12}^{+}\eta_{12}^{-}\left(z_{1}^{2}-1\right)\left(z_{3}^{2}-1\right)\right]
 \\
 \Gamma_{4}&=&\xi_{12}^{-}\eta_{11}^{-}\left[-\xi_{12}^{-}\eta_{11}^{+}z_{1}\left(z_{3}^{2}-1\right)+
 \xi_{11}^{+}\left(\eta_{12}^{-}z_{0}\left(z_{3}^{2}-1\right)+\eta_{12}^{+}\left(z_{1}^{2}+z_{3}^{2}\right)\right)\right]\\\nonumber
 &&\xi_{12}^{+}\xi_{12}^{-}\eta_{11}^{+}\eta_{11}^{-}\left(z_{1}^{2}+1\right)\left(z_{1}^{3}-1\right)-\xi_{12}^{+}
 \xi_{11}^{+}\eta_{11}^{-}\left(\eta_{12}^{-}\left(1+z_{1}^{2}z_{3}^{2}\right)+\eta_{12}^{+}z_{1}\left(z_{1}^{3}-1\right)\right)\\\nonumber
 &&
 +\xi_{12}^{+}\xi_{11}^{-}\eta_{11}^{+}\left[\eta_{12}^{-}\left(z_{1}^{2}+z_{3}^{2}\right)+\eta_{12}^{+}z_{1}\left(1-z_{3}^{2}\right)\right].
\end{eqnarray}

Now we are ready for the computation of the transmission $T_{s}$ and reflection $R_{s}$ 
 probabilities. For this purpose, we introduce
the associated current density $J$, which defines $R_{s}$ and
$T_{s}$ as follows
\begin{equation}
  T_{s}=\frac{ J_{\sf {tra}}}{ J_{\sf {inc}}},\qquad R_{s}=\frac{J_{\sf {ref}}}{ J_{\sf {inc}}}
\end{equation}
where $J_{\sf {\sf {inc}}}$, $J_{\sf {ref}}$ and $J_{\sf {\sf {tra}}}$
stand for the incident, reflected and transmitted components of
the current density, respectively. It is easy to show 
\begin{equation}
J= e\upsilon_{F}\Phi_{\sf }^{\dagger}\sigma _{x}\Phi_{\sf }
\end{equation}
which gives rise to the following results 
\begin{eqnarray}
&& J_{\sf {inc}}=  e\upsilon_{F}(\Phi_{\sf 1}^{+})^{\dagger}\sigma
_{x}\Phi_{\sf 1}^{+}
 \\
 && J_{\sf {ref}}= e\upsilon_{F} (\Phi_{\sf 1}^{-})^{\dagger}\sigma _{x}\Phi_{\sf 1}^{-}
 \\
 && J_{\sf {tra}}= e\upsilon_{F}\Phi_{\sf 5}^{\dagger}\sigma _{x}\Phi_{\sf 5}
\end{eqnarray}
and allow us to express the transmission and reflection
probabilities as
\begin{equation}
  T_{s}=|t_{s}|^{2},
\qquad
  R_{s}=|r_{s}|^{2}.
\end{equation}

We need to point out that we were able to consider a separate transmission probability for each 
of the propagating modes in the leads, because the matching condition does not mix these modes. Under such circumstances, physical measurable quantities such as conductance will be obtained using Landauer formula where summation of over all modes is performed.

\subsection{Numerical results: electrostatic double barrier}
The transmission $T_{s}$ of Dirac electrons in graphene
scattered by a triangular double barrier potential are evaluated
numerically as a function of our system parameters: the
energy $\epsilon$, the $ y $-direction wave vector $ky$, the energy
gap $\mu$, the barriers widths ($d_1,d_2$), the strength of
potential barriers ($v_1, v_2$).

\begin{figure}[ht]
	\centering
	\subfloat[]{
		\centering
		\includegraphics[scale=0.35]{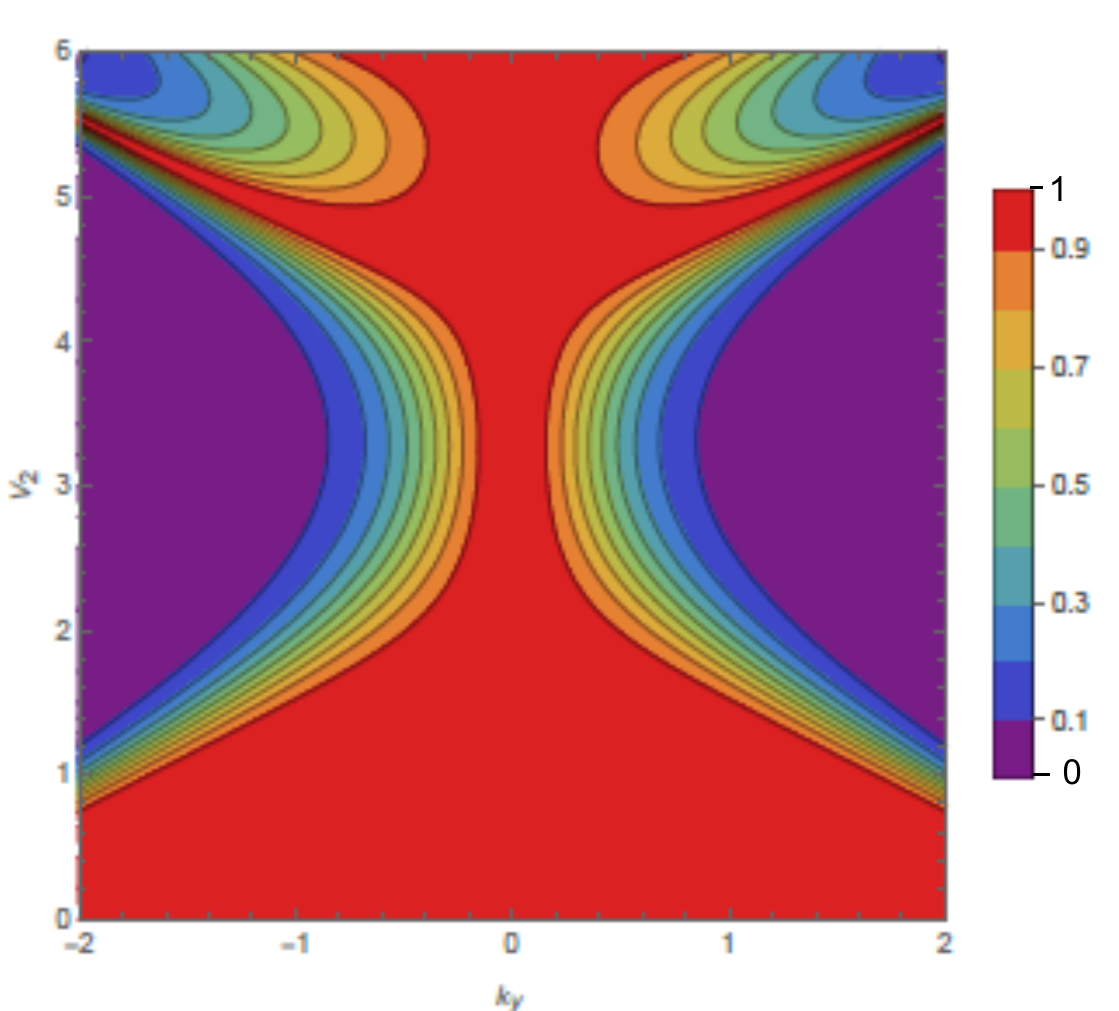}
		\lb{fig2Da}\ \ \ \
	}\subfloat[]{
		\centering
		\includegraphics[scale=0.35]{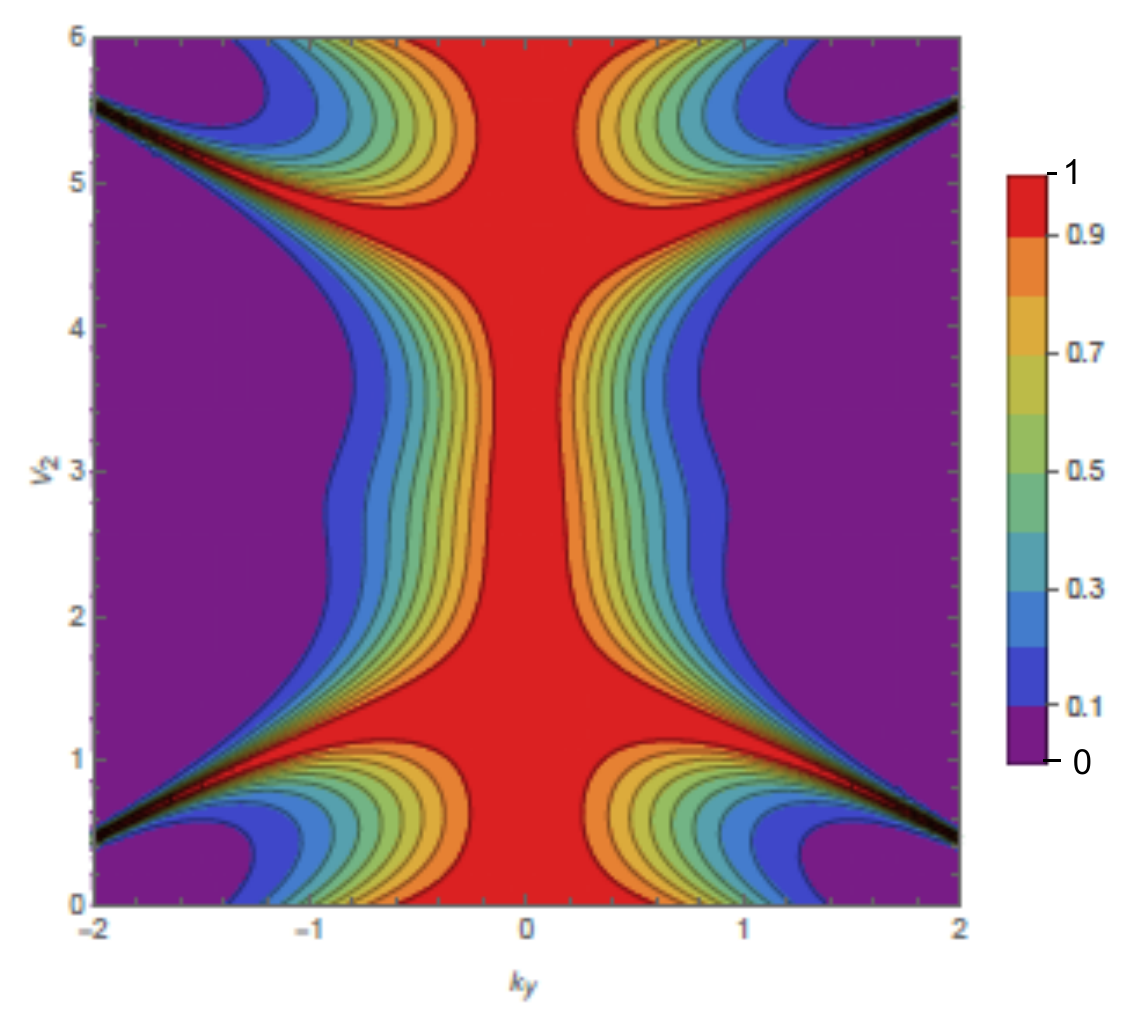}
		\lb{fig2Db}\ \ \ \
	}\subfloat[]{
		\centering
		\includegraphics[scale=0.35]{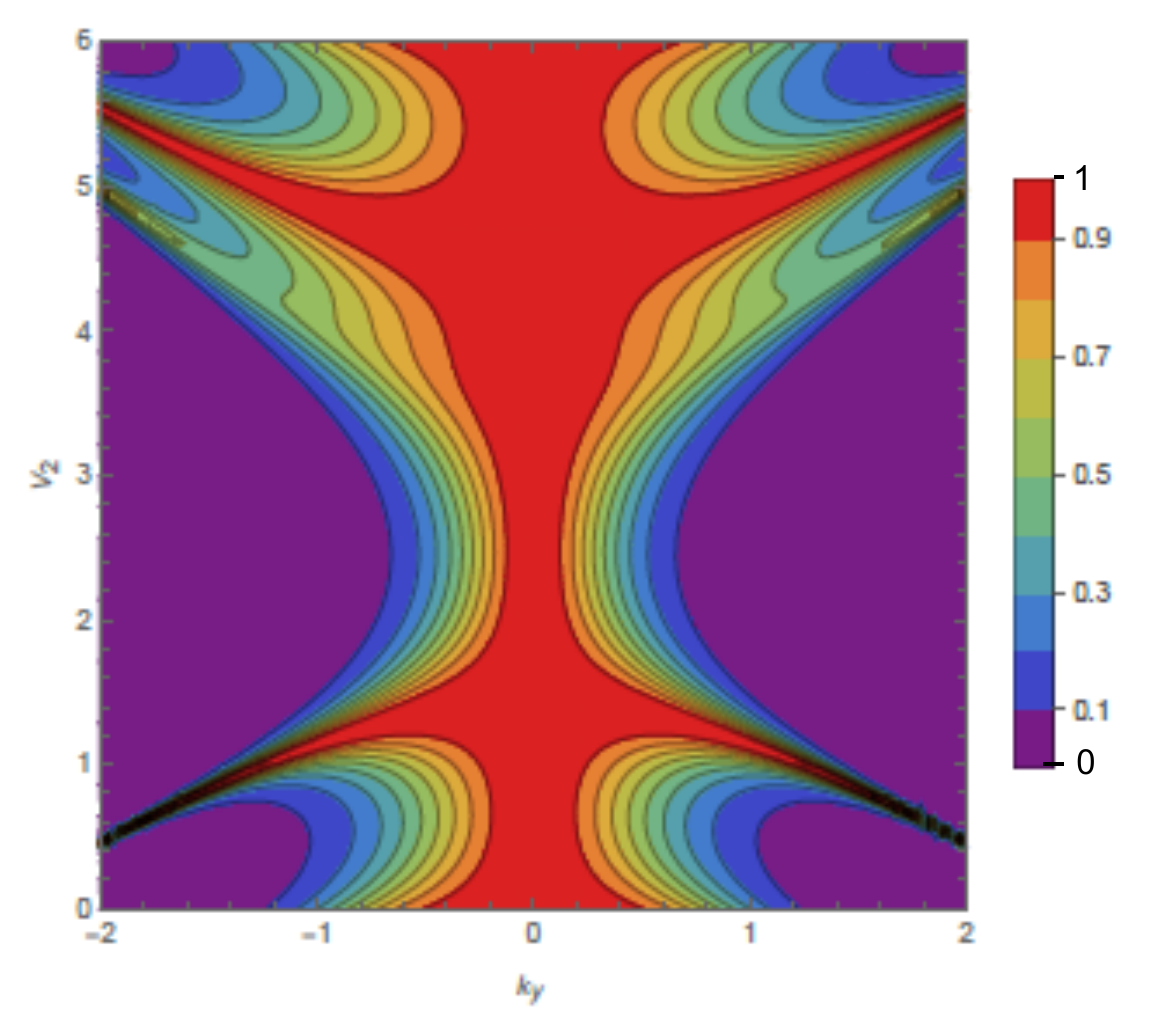}
		\lb{fig2Dc}}
	\caption{\sf{(color online) Contour plot for $T(v_2, k_y)$ with the following parameters:
			$ d_1=1$, $d_2=2$, $\epsilon=3$, $\mu=0$. Let $v_2$ changes from 0
			to 6 and $k_{y}$ from  -2 to -0.001 and 0.001 to 2 that is to say
			$k_{y}$ from  -2 to 2 and  $k_{y}\neq 0$. {\color{red}{(a)}}:
			$v_1=0.5$. {\color{red}{(b)}}: $v_1=4$. {\color{red}{(c)}}:
			$v_1=8$.}}
	\lb{fig005}
\end{figure}

 In Figure \ref{fig005} we
present the contour plot of the transmission probability
$T=tt^{\dagger}$ for a triangular double barrier structure as a
function of well potential $v_2$ and the transverse wave vector $k_y$.
This has been performed by fixing the parameters $d_1=1$, $d_2=2$,
$\epsilon=3$, $\mu=0$ for three different values of the potential height
$v_1=0.5$, $v_1=4$ and $v_1=8$ corresponding to figures
\ref{fig2Da}, \ref{fig2Db} and \ref{fig2Dc}, respectively. Note that the strength
of the potential height $v_1$ in this figure is proportional to the strength of the applied electrostatic field. The different colors from purple to red correspond to different values
of the transmission continuously from $0$ to $1$. The number of oscillations
increases as the strength of the electric field increases. The resonances that
are clear in the transmission probability show up as peaks, we also
observe the occurrence of a new strong and sharp resonance around $v_2=\epsilon$.

\begin{figure}[ht]
        \centering
        \subfloat[]{
            \centering
            \includegraphics[scale=0.35]{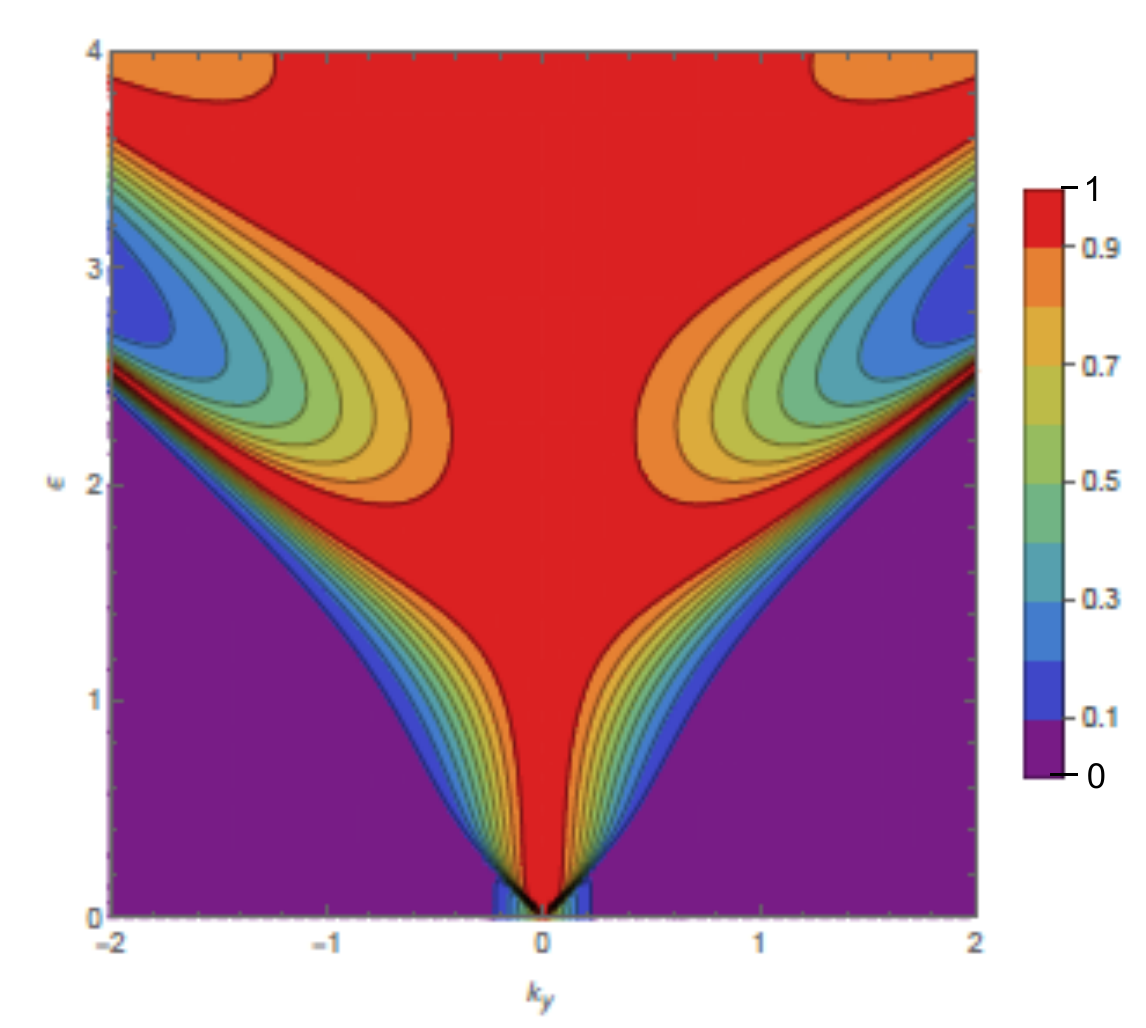}
            \lb{fig3Da}\ \ \ \
        }\subfloat[]{
            \centering
            \includegraphics[scale=0.35]{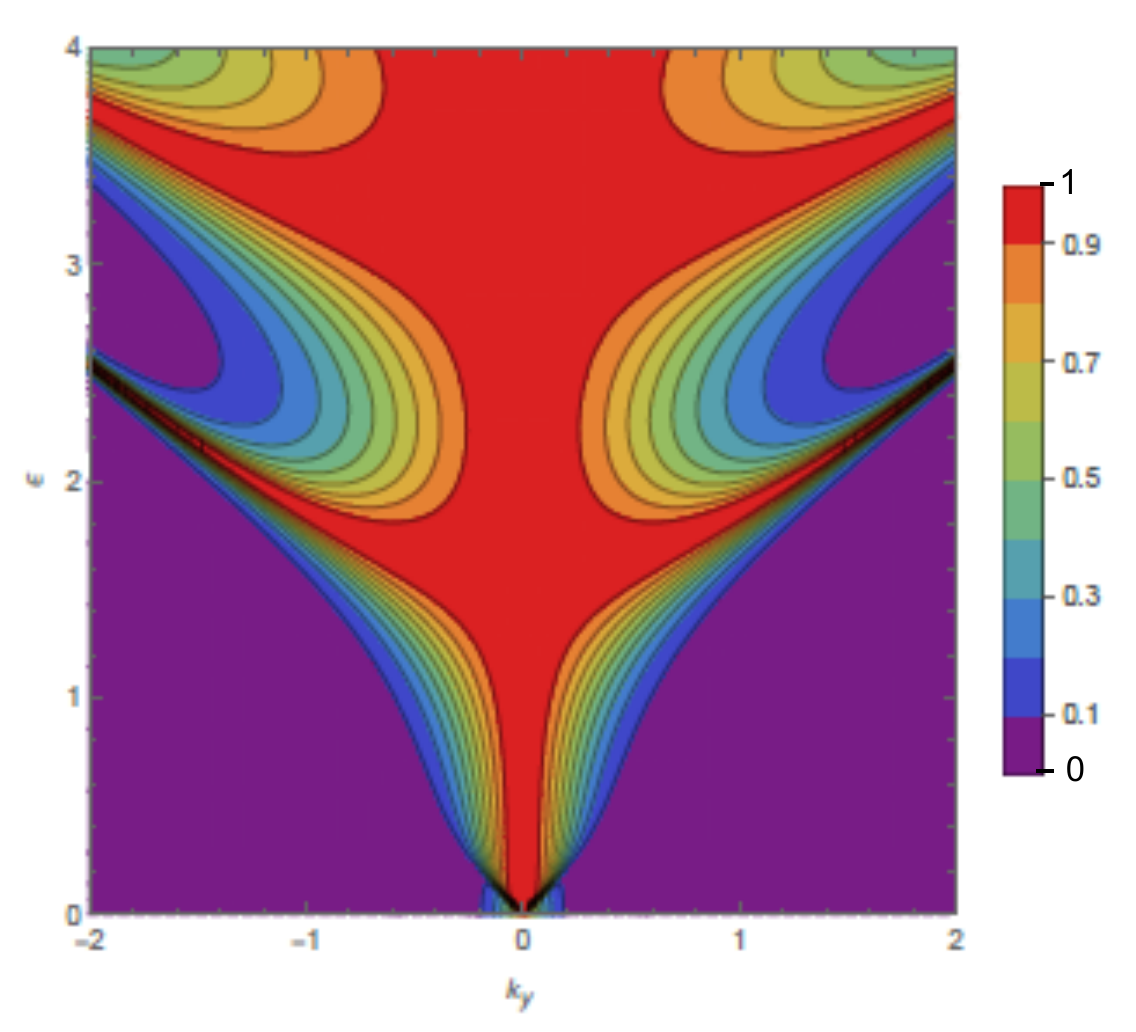}
            \lb{fig3Db}\ \ \ \
            }\subfloat[]{
            \centering
            \includegraphics[scale=0.35]{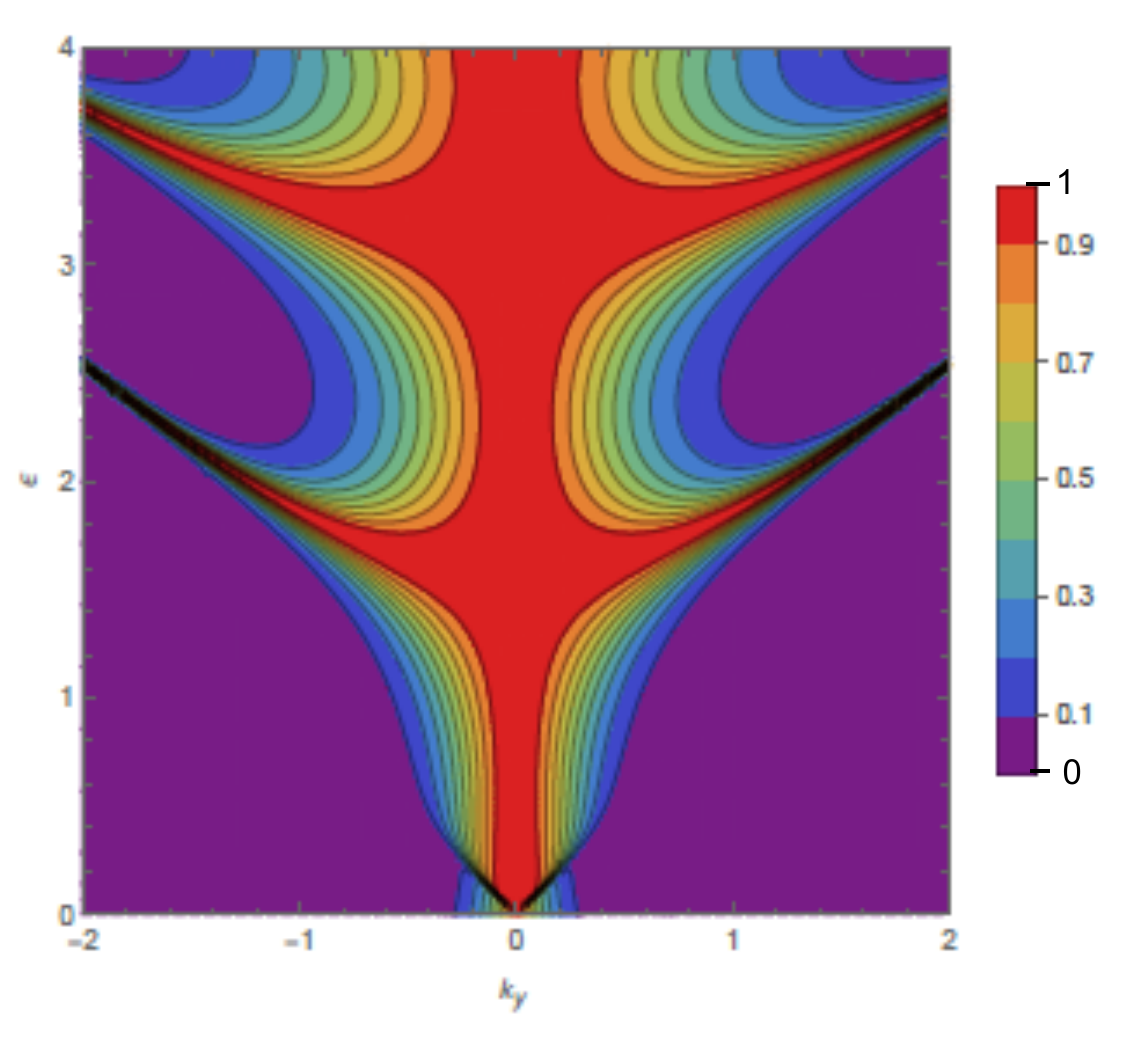}
            \lb{fig3Dc}}
        \caption{\sf{(color online) Contour plot for $T(\epsilon, k_y)$ with the following parameters:
$ d_1=1$, $d_2=2$, $\mu=0$, $v_2=0$. Let $\epsilon$ changes from 0
to 4 and $k_{y}$ from  -2 to -0.001 and then from 0.001 to 2 that is to say
$k_{y}$ from  -2 to 2 and  $k_{y}\neq 0$. {\color{red}{(a)}}:
$v_1=2$. {\color{red}{(b)}}: $v_1=3$. {\color{red}{(c)}}:
$v_1=5$.}}
      \lb{fig008}
    \end{figure}
    
    {{ In Figure \ref{fig008} we show the contour plot of the transmission
    		probability as function of energy $\epsilon$ and the transfer wave
    		vector $k_y$ for $ d_1=1$, $d_2=2$, gapless graphene $\mu=0$
    		and  three different values for potential $v_1$. Here, we take the
    		case with $v_1=2$ in figure \ref{fig3Da} initially and then increase
    		$v_1$ from $v_1=3$ in figure \ref{fig3Db} to $v_1=5$ in
    		figure \ref{fig3Dc} and notice that the number of oscillations increases, when the
    		incidence is almost normal, the wave vector is close to $k_y\approx 0$
    		associated with total transmission. We see that full transmission is squeezed into a very narrow angle region around normal incidence, $k_y\approx 0$, when we increase the strength of the electrostatic field make it favorable application to tunneling collimation.}}
    	
    \begin{figure}[ht]
    	\centering
    	\subfloat[]{
    		\centering
    		\includegraphics[scale=0.35]{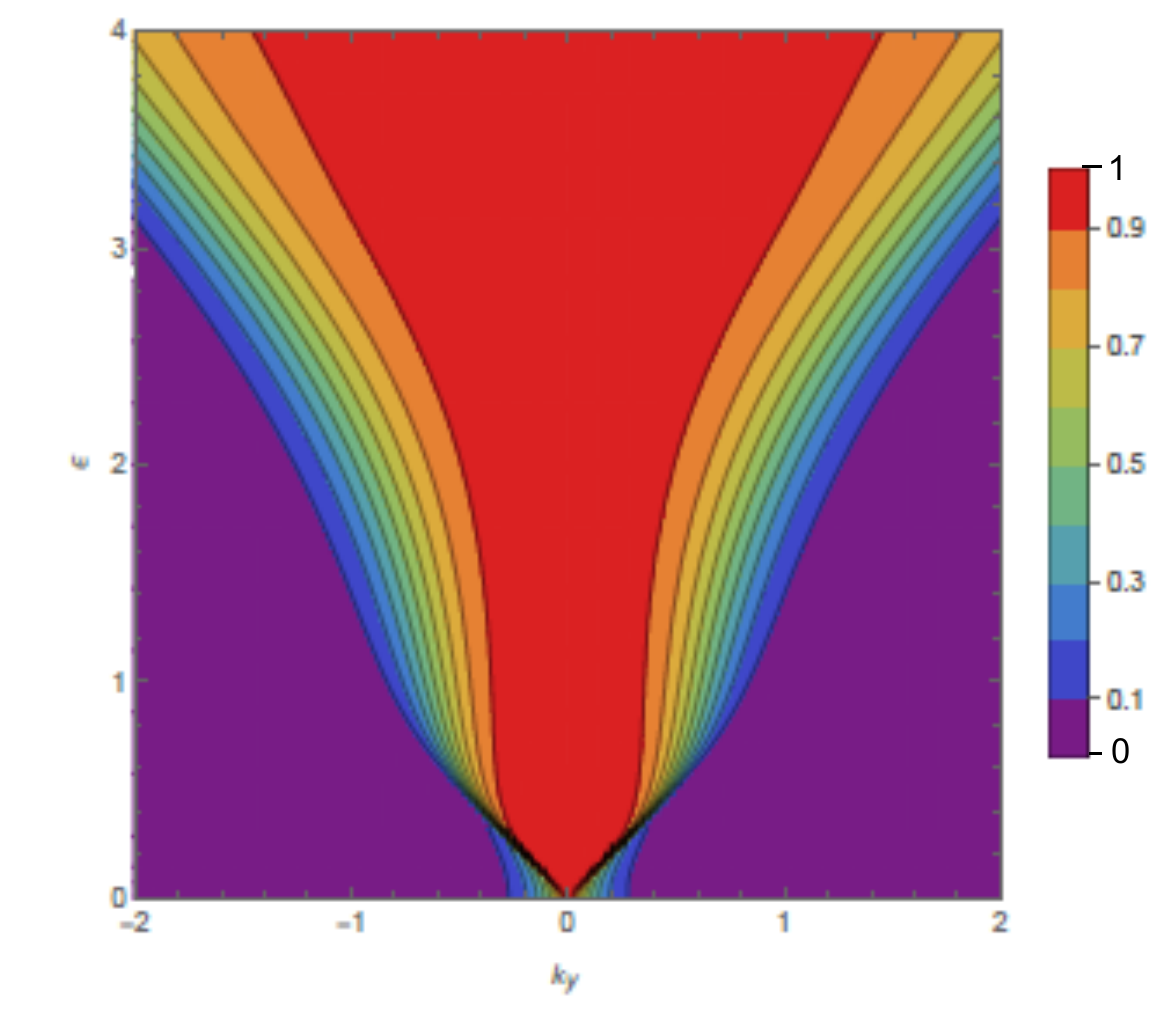}
    		\lb{fig1Da}\ \ \ \
    	}\subfloat[]{
    		\centering
    		\includegraphics[scale=0.35]{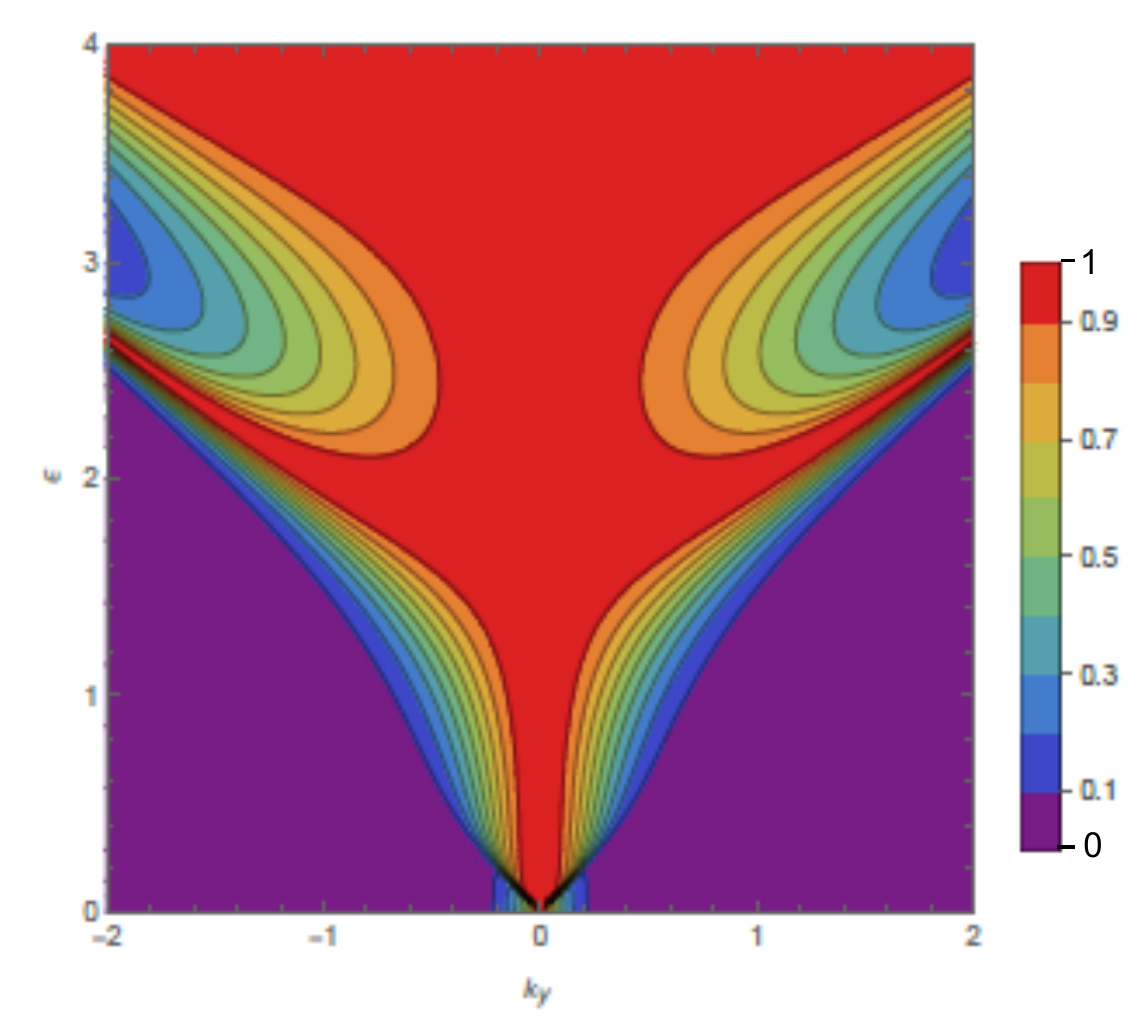}
    		\lb{fig1Db}\ \ \ \
    	}\subfloat[]{
    		\centering
    		\includegraphics[scale=0.35]{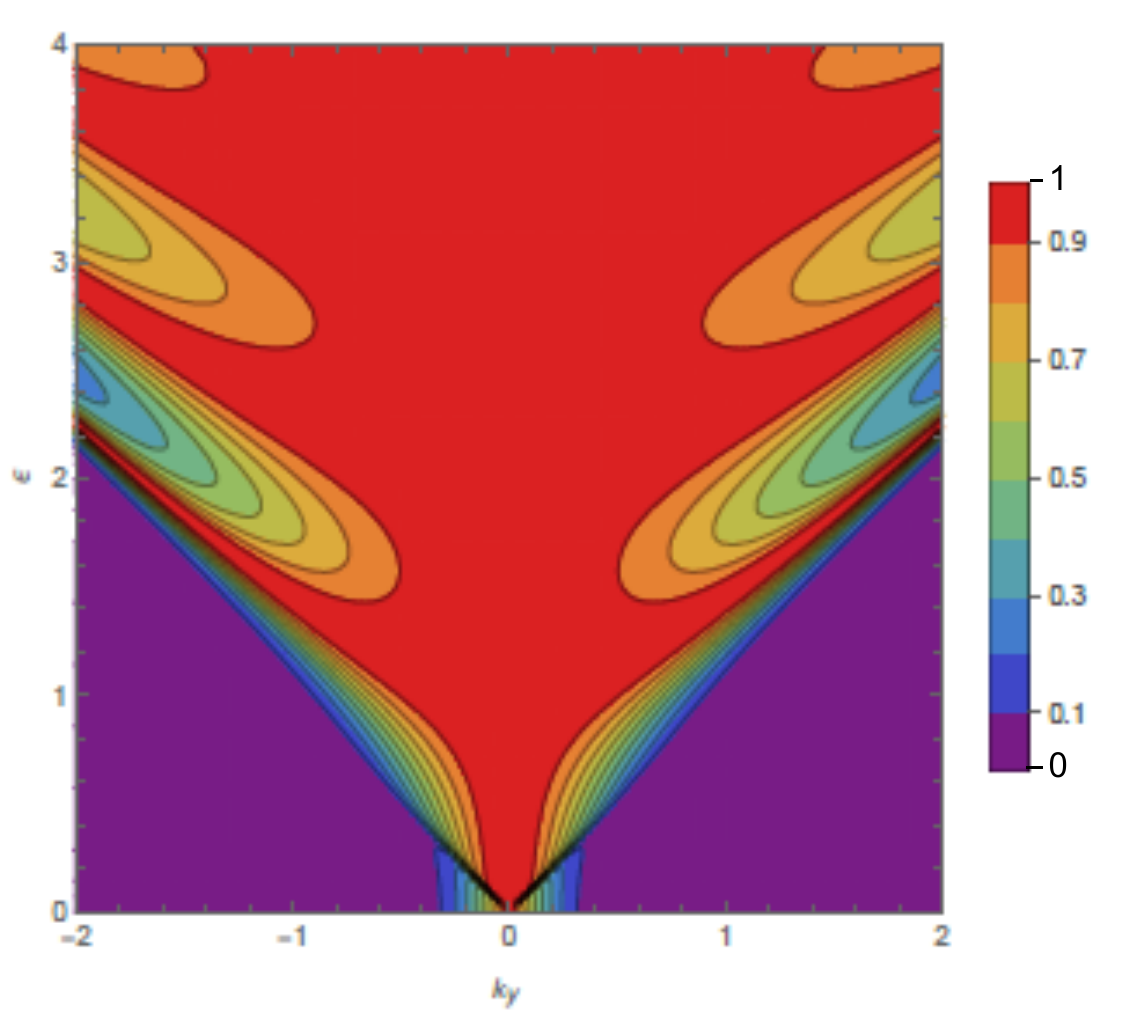}
    		\lb{fig1Dc}}
    	\caption{\sf{(color online) Contour plot for $T(\epsilon, k_y)$ with the following parameters:
    			$ v_1=2$, $d_2=2$, $\mu=0$, $v_2=0$. Let $\epsilon$ changes from 0
    			to 4 and $k_{y}$ from  -2 to -0.001 and then from 0.001 to 2 that is to say
    			$k_{y}$ from  -2 to 2 and  $k_{y}\neq 0$. {\color{red}{(a)}}:
    			$d_1=0.1$. {\color{red}{(b)}}: $d_1=0.9$. {\color{red}{(c)}}:
    			$d_1=1.5$.}}
    	\lb{fig007}
    \end{figure}

 {{ In Figure \ref{fig007} we show the contour plot of the transmission
probability as function of energy $\epsilon$ and the transverse wave
vector $k_y$ from -2 to 2 and  $k_{y}\neq 0$, we fix the potential
$ v_1=2$,  $v_2=0$, the barrier width $d_2=2$, for gapless
graphene $\mu=0$ and three different values of the barrier width
$d_1=0.1$ in figure \ref{fig1Da}, $d_1=0.0.9$ in figure
\ref{fig1Db} and $d_1=1.5$ in figure \ref{fig1Dc}. By increasing
the width $d_1$ the number of oscillation peaks increases.
The parameters of the triangular potential ($v_1$, $d_2-d_1$), whose ratio represents the strength of the applied electrostatic field,  are
found to play a key role in controlling the tunneling
resistance peak. All these predicted attractive transport properties are expected to be
extremely useful for designing both novel electronic and optical
graphene-based devices and electronic lenses in ballistic-electron
optics.}}

\begin{figure}[ht]
	\centering
	\subfloat[]{
		\centering
		\includegraphics[scale=0.495]{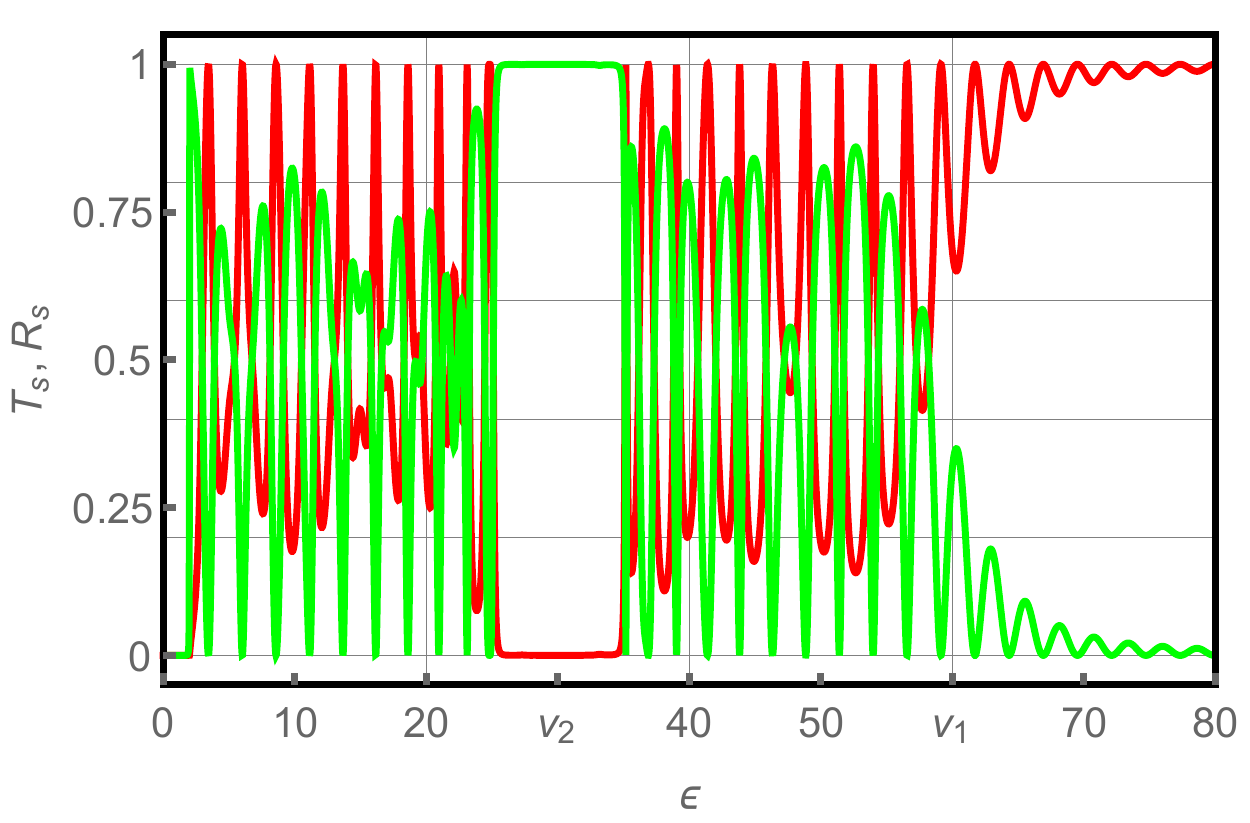}
		\label{fig11a}\ \ \ \
	}\subfloat[]{
		\centering
		\includegraphics[scale=0.49]{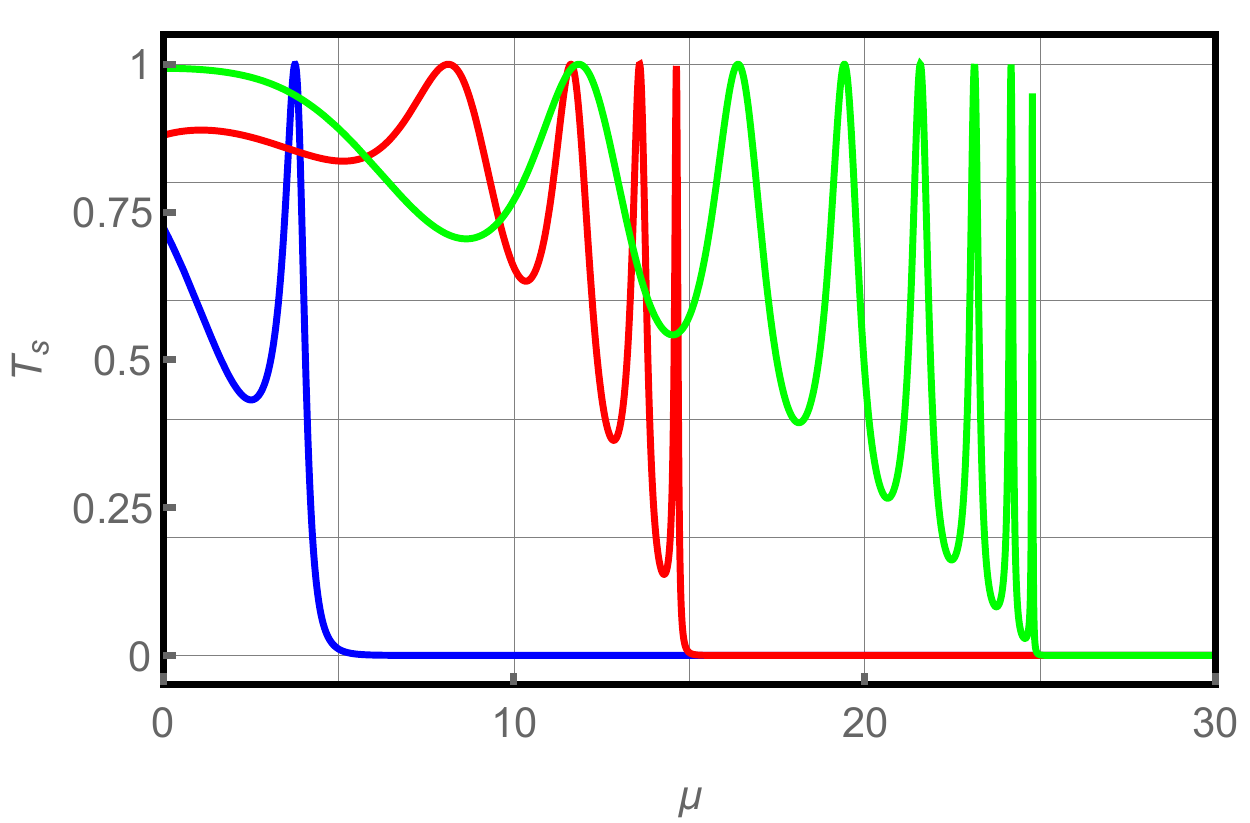}
		\label{fig11b}}
	\caption{(color online) \sf{{\color{red}{(a)}}: Transmission (red color) and reflection (green color) probabilities $(T_{s}, R_{s})$ as a function of energy
			$\epsilon$ with $d_{1}=0.6$, $d_{2}=2.5$, $\mu=4$,
			$k_{y}=2$,
			$v_{1}=60$ and $v_{2}=30$. {\color{red}{(b)}}: Transmission probability $T_{s}$ as a function of
			energy gap $\mu$ with $d_{1}=0.5$, $d_{2}=1.5$,
			$k_{y}=1$,
			$v_{1}=50$, $v_{2}=40$, {{$\epsilon=15$ (green color), $\epsilon=25$ (red color) and $\epsilon=35$ (blue color)}}.}}
	\lb{fig1ab}
\end{figure}

   {{ In Figure \ref{fig11a} we show the transmission
and reflection probabilities versus the energy $\epsilon$.
Obviously, we can check that the probability conservation
condition $T_{s}+R_{s}=1$ is well satisfied. In the first energy
interval $\epsilon \leq k_{y}$ we have no transmission because it
is a forbidden zone.
However, in the second energy intervals $k_{y} \leq \epsilon \leq
v_{2}-k_{y}-\frac{\mu}{2}$ and
$v_{2}+k_{y}+\frac{\mu}{2}\leq\epsilon\leq v_{1}$, we observe
resonance oscillations due to the Klein regime. We have no
transmission (gap region) when
$v_{2}-k_{y}-\frac{\mu}{2}\leq\epsilon\leq
v_{2}+k_{y}+\frac{\mu}{2}$. Finally, in the interval where
$\epsilon > v_{1}$, there exist usual high energy oscillations,
which saturate asymptotically. Note that
\eqref{1313} implies that for certain energy gap $\mu$, there is
no transmission. In fact, under the condition
$
\mu>|v_{2}-\epsilon|
$
every incoming wave is reflected. In Figure \ref{fig11b}  we see
that the transmission vanishes for values of $\epsilon$ below the
critical value
$\mu=|v_{2}-\epsilon|$}}.

\begin{figure}[ht]
	\centering
	\subfloat[]{
		\centering
		\includegraphics[scale=0.5]{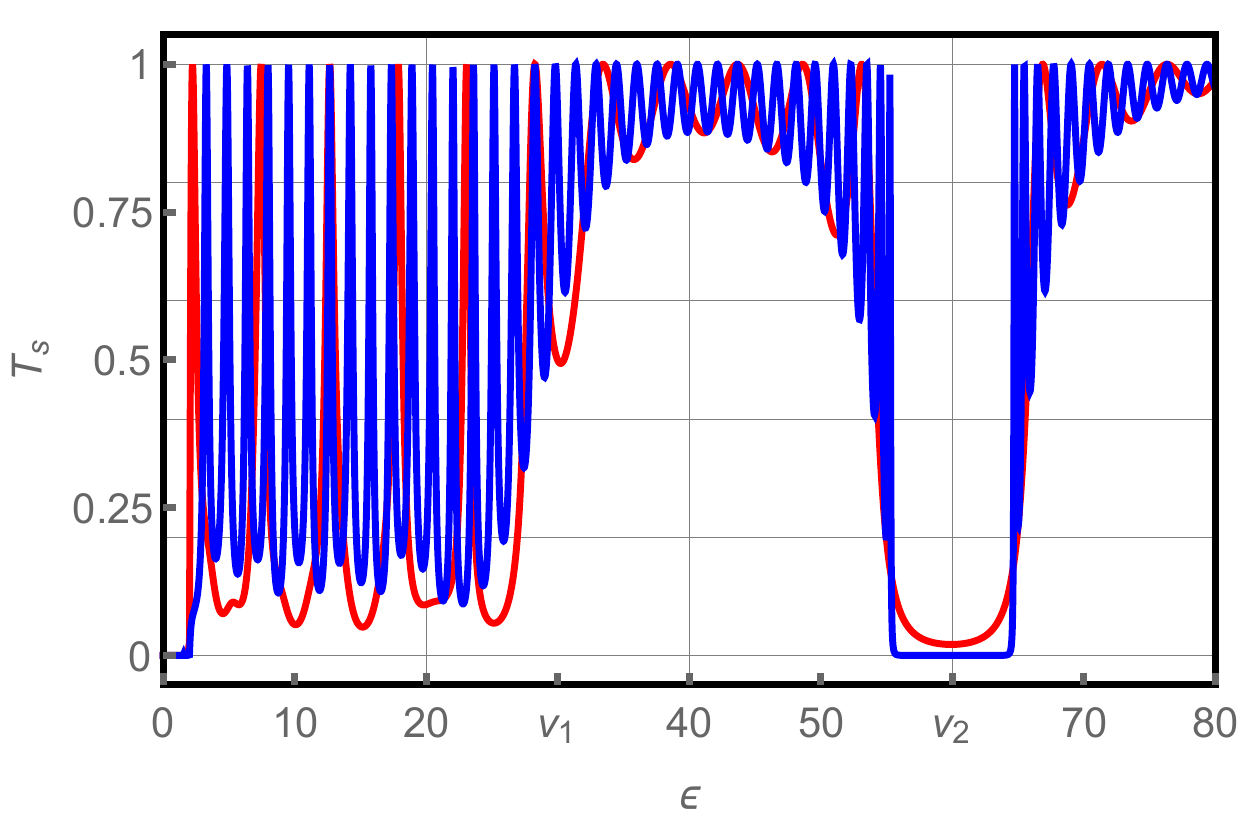}
		\label{fig22a}\ \ \ \
	}\subfloat[]{
		\centering
		\includegraphics[scale=0.5]{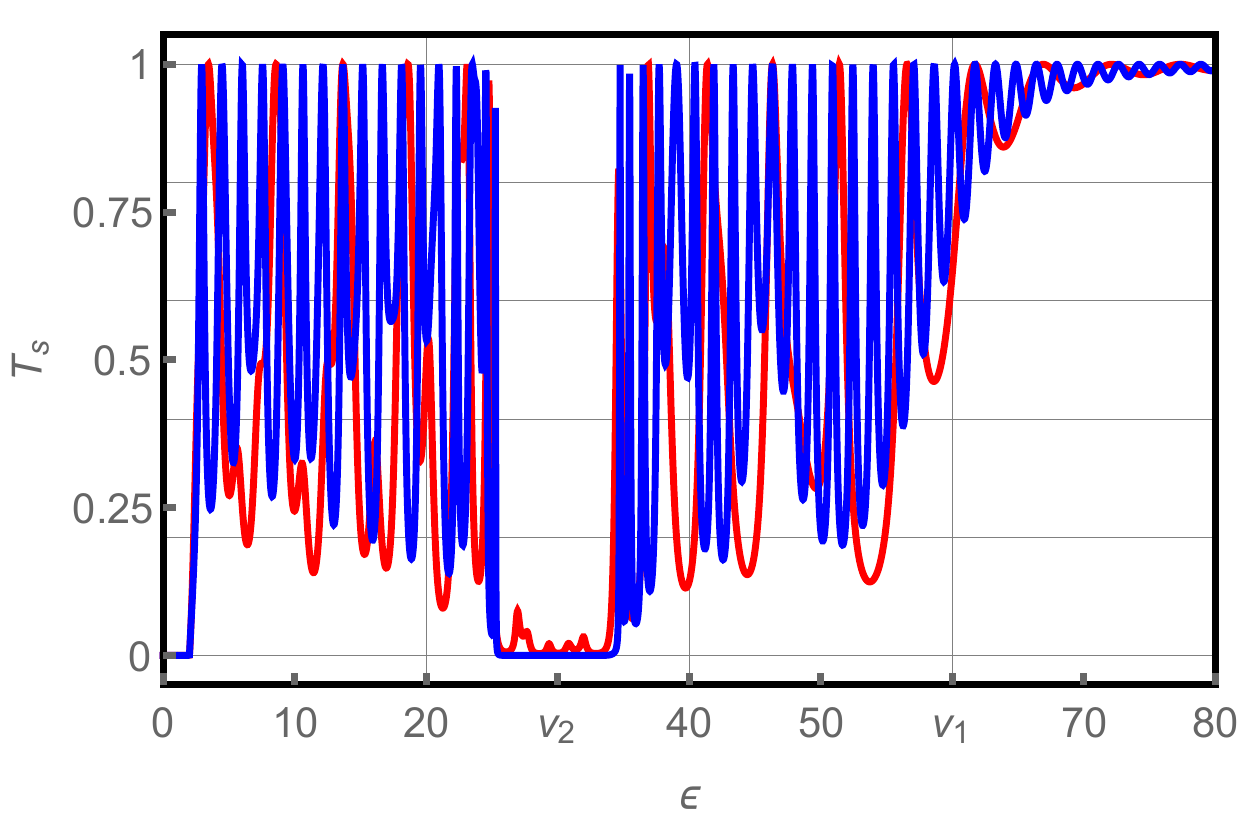}
		\label{fig22b}}
	\caption{\sf{(color online) Transmission probability for the static barrier $T_{s}$ as a function of energy
			$\epsilon$ with  $d_{1}=0.3$ (red color), $d_{1}=1$ (blue color), $d_{2}=2.5$, $\mu=4$ and $k_{y}=2$. {\color{red}{(a)}}:  {{for $v_{1} \leq v_{2}$: $v_{1}=30$, $v_{2}=60$}}.
			{\color{red}{(b)}}: {{for $v_{1} \geq v_{2}$:
					$v_{1}=60$, $v_{2}=30$}}.}}
	\lb{figgi3}
\end{figure}

Figure \ref{figgi3} shows the transmission $T_{s}$ as a
function of incident electron energy $\epsilon$ for the Dirac
fermion scattered by a double triangular barriers with
$d_{2}=2.5$, $\mu=4$, $k_{y}=2$ and two values of barrier width
$d_{1}=\{0.3, 1\}$. In Figure \ref{fig22a} we
consider the parameters $v_{1}=\frac{v_{2}}{2}=30$ for the Dirac
fermion scattered by a double barrier triangular potential where
we distinguish five  different zones.
Indeed,
the first is a forbidden zone where 0$ \leq \epsilon \leq k_{y}$.
 The
second zone $k_{y} \leq \epsilon \leq v_{1}$ is the lower Klein
energy zone with transmission resonances.
 The third zone contains transmission oscillations.
 The fourth one
$v_{2}-k_{y}-\frac{\mu}{2}\leq\epsilon\leq
v_{2}+k_{y}+\frac{\mu}{2}$ is a window where the transmission is
zero, the wavefunction is damped and transmission decays
exponentially.
The fifth zone  $\epsilon \geq
v_{2}+k_{y}+\frac{\mu}{2}$ contains over barrier oscillations, the transmission
converges to unity at high energies similarly to the non-relativistic result.
We also notice from Figure \ref{fig22a} that resonances move to  the right as we
increase the width of the field region (reduction of the electrostatic field).
In Figure \ref{fig22b} we consider the opposite situation, that is
$v_{1}=2 v_{2}=60$, the results show that as long as the well width
$d_{1}$ increases the transmission resonances shift  to the left and the width
of the resonances increase in regions $k_{y} \leq \epsilon \leq
v_{2}-k_{y}-\frac{\mu}{2}$ and $v_{2} + k_{y} +
\frac{\mu}{2}\leq\epsilon\leq v_{1}$.

\begin{figure}[ht]
        \centering
        \subfloat[]{
            \centering
            \includegraphics[scale=0.51]{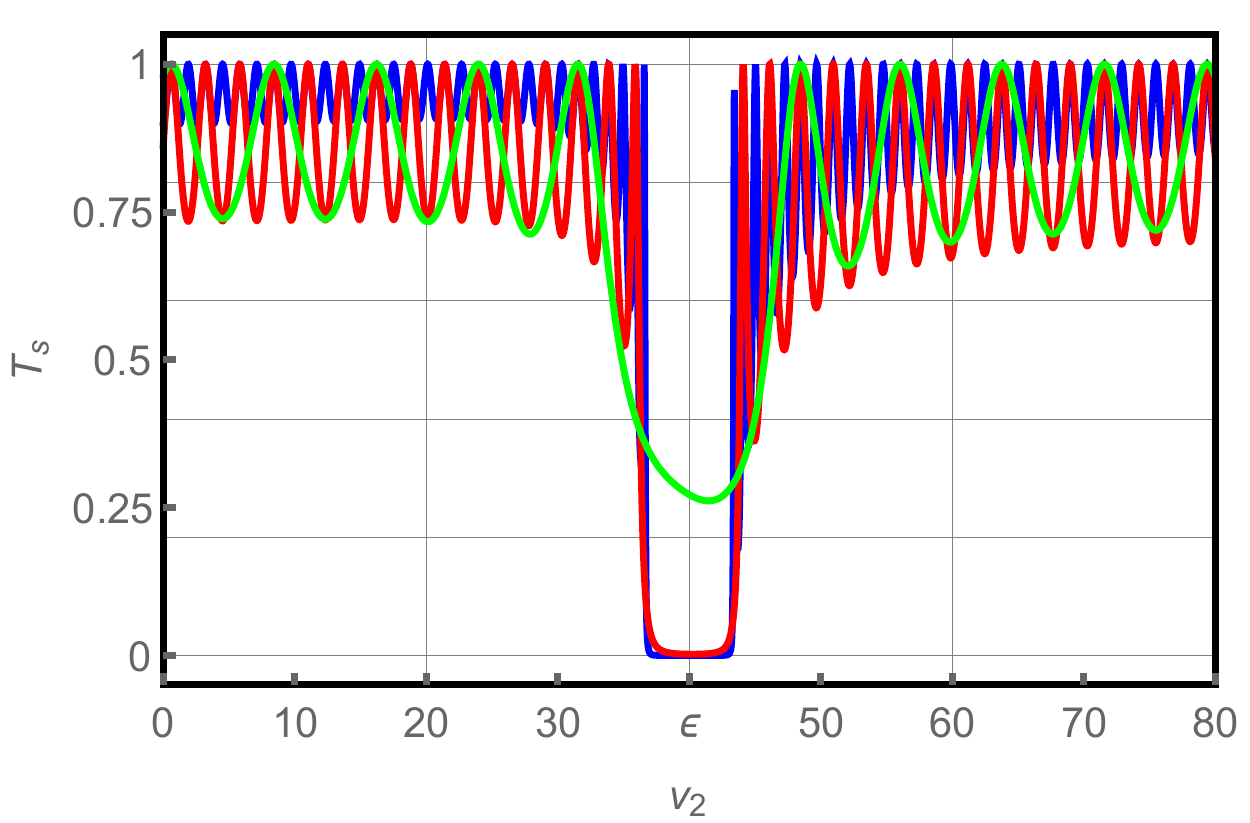}
            \lb{figg4a}
        }\subfloat[]{
            \centering
            \includegraphics[scale=0.51]{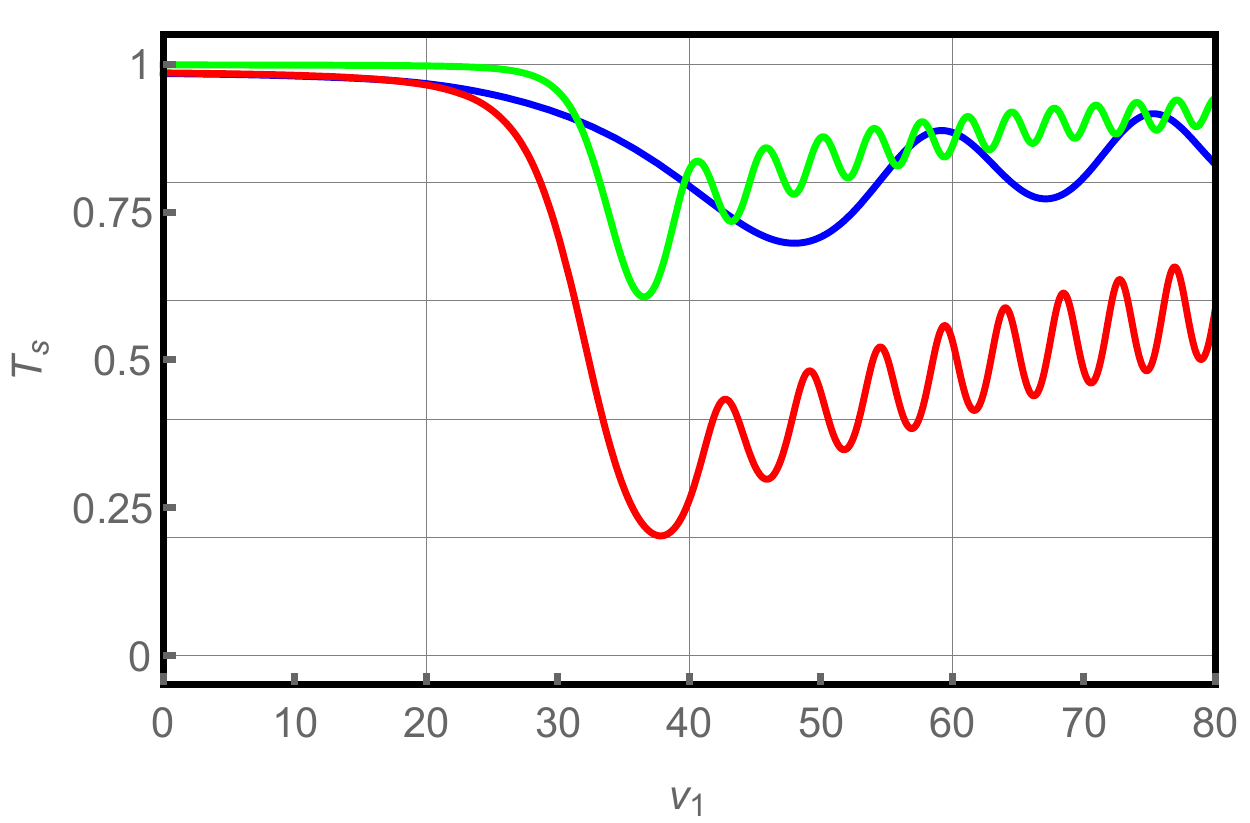}
            \lb{figg4b}}
        \caption{\sf{(color online) {\color{red}{(a)}}: Transmission probability through the static barrier $T_{s}$ as a function of the potential
 $v_{2}$ for {{$d_{1}=0.2$ (green color), $d_{1}=0.6$ (red color), $d_{1}=1.2$ (blue color)}}, $d_{2}=2$, $\mu=3$, $k_{y}=1$,
 $\epsilon=40$ and $v_{1}=60$. {\color{red}{(b)}}: Transmission probability  through the static barrier $T_{s}$ as a function of the potential
 $v_{1}$ with {{$d_{1}=0.05$ (green color), $d_{1}=0.7$ (red color), $d_{1}=2$ (blue color)}}, $d_{2}=2.5$, $\mu=4$, $k_{y}=2$,
 $\epsilon=30$ and $v_{2}=60$.}}
      \lb{figg4}
    \end{figure}
We present in Figure \ref{figg4a} the transmission versus the
potential $v_{2}$. It is clear that the two transmission curves are
almost symmetric with respect to the point $v_{2} = \epsilon$. While an
increase in the value $d_{1}$ widens the bowl width. As we increase the well width
we see an increase in the number of resonances while a zero transmission region appears
only for small values of $d_{1}$ (large values of the electrostatic field). Figure \ref{figg4b} shows the transmission probability  for a static barrier $T_{s}$  as function of the strength of the applied
voltage $v_{1}$ for different widths of the electric field region. Full transmission is observed for small values of $v_{1}$, less than the energy of the incident fermion. It then
decreases sharply for $v_{1} > \epsilon -(2k_{y}+\mu)$ until it reaches a relative minimum and then begins to increase in an oscillatory manner. However, we noticed that a gap region appeared for energies around the value of the elevated well potential, $v_{2}-k_{y}-\frac{\mu}{2}\leq\epsilon\leq
v_{2}+k_{y}+\frac{\mu}{2}$, independently of the ordering strength of the potentials $v_{1}$ and $v_{2}$. As a results, we observe that 
the double barrier strongly affect the transmission 
and then it can be used a tunable parameter to control
the tunneling in  our system, which can be used in technological applications. 

\section{Magnetic double barrier}
Consider a two-dimensional system of Dirac fermions forming a
graphene sheet. This sheet is subject to a double barrier
potential in addition to a mass term and an externally applied
magnetic field. Particles and antiparticles moving in
the positive and negative energy regions, respectively, have a conserved
component of the wave vector along the $y$-direction (mode) due to
the fact that each of the $ N $ propagating modes in the leads is independent because
the matching condition does not mix these modes.
%
A uniform perpendicular
magnetic field is applied, along the $z$-direction and confined to the well
region between the two electrostatic barriers. It is defined by
\begin{equation}\label{eq04}
B(x,y)=B\Theta(d_{1}^{2}-x^{2})
\end{equation}
where $B$ is the strength of the magnetic field within the strip
located in the region $|x|< d_{1}$ and $B=0$ otherwise, $\Theta$ is
the Heaviside unit step function. Choosing the Landau gauge and
imposing continuity of the vector potential at the boundary to
avoid unphysical effects, we end up with the following vector
potential
\begin{equation}
\qquad A_{y}(x)=A_{j}=\frac{c}{e}\times\left\{%
\begin{array}{ll}
  -\frac{1}{l_{B}^{2}}d_{1}, & \hbox{$x<-d_{2}$} \\
    \frac{1}{l_{B}^{2}}x, & \hbox{$\mid x\mid<d_{1}$} \\
     \frac{1}{l_{B}^{2}}d_{1}, & \hbox{$x\geq d_{2}$} \\
\end{array}%
\right.
\end{equation}
with the magnetic field length $l_{B}=\sqrt{1/B}$ in the unit
system ($\hbar=c=e=1$). 
The system is composed of five regions denoted ${\sf j=1,2,3,4,5}$. The
left region (${\sf j=1}$) describes the incident electron beam
with energy $E=v_{F}\epsilon$ at an incident angle
$\phi_{1}$  where $v_{F}$ is the Fermi velocity. The extreme right region
(${\sf j=5}$) describes the transmitted electron beam at an
angle $\phi_{5}$. The Hamiltonian for describing our system reads as
\begin{equation} \lb{equm1}
H_{m}=v_{F}
{\boldsymbol{\sigma}}\cdot\left(\textbf{p}+\frac{e}{c}\textbf{A}\right)+
V(x){\mathbb I}_{2}+G_p\Theta\left(d_{1}^{2}-x^{2}\right)\sigma_{z}.
\end{equation}
{To proceed further,} we need to find the solutions of
the corresponding Dirac equation and their associated energy
spectrum.

\subsection{Energy spectrum: presence of magnetic field}
We are set to determine the eigenvalues and eigenspinors of the Hamiltonian
 $H_{m}$. Indeed, the Dirac Hamiltonian describing regions 1 and 5, is
obtained from \eqref{equm1} as
\begin{equation}
H_{m}=\left(%
\begin{array}{cc}
  0 & \upsilon_{F}\left(p_{x{\sf j}}-i\left(p_{y}+\frac{e}{c}A_{\sf j}\right)\right) \\
  \upsilon_{F}\left(p_{x{\sf j}}+i\left(p_{y}+\frac{e}{c}A_{\sf j}\right)\right) & 0 \\
\end{array}%
\right).
\end{equation}
The corresponding time independent Dirac equation for the spinor
$\psi_{\sf j}(x,y)= (\varphi_{\sf j}^{+}, \varphi_{\sf j}^{-})^{T}$ at energy
$E=\upsilon_{F}\epsilon$ is given by
\begin{equation}
H_{m}\left(%
\begin{array}{c}
  \varphi_{\sf j}^{+} \\
  \varphi_{\sf j}^{-} \\
\end{array}%
\right)=\epsilon\left(%
\begin{array}{c}
  \varphi_{\sf j}^{+} \\
  \varphi_{\sf j}^{-} \\
\end{array}%
\right).
\end{equation}
This eigenvalue problem can be written as two linear differential
equations of the from
\begin{eqnarray}
     &&p_{x{\sf j}}-i\left(p_{y}+\frac{e}{c}A_{\sf j}\right)\varphi_{\sf j}^{-}=\epsilon\varphi_{\sf j}^{+}  \\
     && p_{x{\sf j}}+i\left(p_{y}+\frac{e}{c}A_{\sf j}\right)\varphi_{\sf j}^{+}=\epsilon\varphi_{\sf j}^{-}
\end{eqnarray}
which gives the energy eigenvalue
\begin{equation}
\epsilon=s_{\sf j} \sqrt{p_{x{\sf j}}^{2}+\left(p_{y}+\frac{e}{c}A_{\sf j}\right)^{2}}
\end{equation}
where $s_{\sf j}=\mbox{sign}(\epsilon)$,
 the incoming momentum ${\boldsymbol{p_{\sf j}}}=(p_{x{\sf j}}, p_{y})$ and
position ${\boldsymbol{r}}=(x, y)$. The incoming spinor is
\begin{eqnarray}
 \psi_{in}=\frac{1}{\sqrt{2}}\left(
\begin{array}{c}
1 \\
 z_{p_{x{\sf j}}}\end{array}\right)e^{\textbf{\emph{i}}{\boldsymbol{p_{\sf j}}}\cdot {\boldsymbol{r}}}, \qquad
 z_{p_{x{\sf j}}}=z_{\sf j}=
s_{\sf j}
e^{\textbf{\emph{i}}\phi_{\sf j}}
\end{eqnarray}
with  $s_{0}=\mbox{sgn}(\epsilon)$ and  $\phi_{\sf j}=\arctan\left(\frac{p_{y}-\frac{e}{c}A_{\sf j}}{p_{x{\sf j}}}\right)$ is the angle that
the incident electron beam makes with the {$x$-direction}, $p_{x1}$ and
$p_{y}$ are the $x$ and $y$-components of the electron wave
vector, respectively. The eigenspinors are given by
\begin{eqnarray}
&& \psi_{\sf j}^{+}=\frac{1}{\sqrt{2}}\left(
\begin{array}{c}
1 \\
 z_{\sf j}\end{array}\right)e^{\textbf{\emph{i}}(p_{x{\sf j}} x +p_{y} y)}\\
&&
\psi_{\sf j}^{-}=\frac{1}{\sqrt{2}}\left(
\begin{array}{c}
1 \\
 -z^{*}_{\sf j}\end{array}\right)e^{\textbf{\emph{i}}(-p_{x{\sf j}} x +p_{y} y)}.
\end{eqnarray}
It is straightforward to solve the tunneling problem for Dirac
fermions. We assume that the incident  wave propagates at the angle
$\phi_{1}$ with respect to the {$x$-direction} and write  the
components, of the Dirac spinor $\varphi_{\sf j}^{+}$ and
$\varphi_{\sf j}^{-}$, for each
region. 
For $x<-d_{2}$ (region {\sf 1}):
\begin{eqnarray}
&& \epsilon=
\left[p_{x1}^{2}+\left(p_{y}-\frac{1}{l_{B}^{2}}d_{1}\right)^{2}\right]^{\frac{1}{2}}\\
&& \psi_{1}=\frac{1}{\sqrt{2}}\left(
\begin{array}{c}
1 \\
 z_{1}\end{array}\right)e^{\textbf{\emph{i}}(p_{x1} x +p_{y} y)}+r_{m}\frac{1}{\sqrt{2}}\left(
\begin{array}{c}
1 \\
 -z^{*}_{1}\end{array}\right)e^{\textbf{\emph{i}}(-p_{1x} x +p_{y}
 y)}\\
&& z_{1}=s_{1}\frac{p_{x1}
+i\left[p_{y}-\frac{1}{l_{B}^{2}}d_{1}\right]}{\sqrt{p_{x1}^{2}
+\left[p_{y}-\frac{1}{l_{B}^{2}}d_{1}\right]^{2}}}.
\end{eqnarray}

In the barrier $x
> d_{2}$ (region {\sf 5}):
\begin{eqnarray}
&& \epsilon=\left[p_{x5}^{2}+\left(p_{y}+\frac{1}{l_{B}^{2}}d_{1}\right)^{2}\right]^{\frac{1}{2}}\\
&& \psi_{5}=\frac{1}{\sqrt{2}}t_{m}\left(
\begin{array}{c}
1 \\
 z_{5}\end{array}\right)e^{\textbf{\emph{i}}(p_{x5} x +p_{y} y)}\\
&& z_{5}=s_{5}\frac{p_{x5}
+i\left[p_{y}+\frac{1}{l_{B}^{2}}d_{1}\right]}{\sqrt{p_{x1}^{2}
+\left[p_{y}+\frac{1}{l_{B}^{2}}d_{1}\right]^{2}}}.
\end{eqnarray}

In region {\sf 2} and {\sf 4} ($d_{1}<|x|<d_{2}$):
The general solution can be expressed in terms of the parabolic
cylinder function \cite{Abramowitz, Gonzalez, HBahlouli} as
\begin{equation}\lb{hii1}
 \chi_{\gamma}^{+}=c_{1}
 D_{\nu_\gamma-1}\left(Q_{\gamma}\right)+c_{2}
 D_{-\nu_\gamma}\left(-Q^{*}_{\gamma}\right)
\end{equation}
where
$\nu_{\gamma}=\frac{i}{2\varrho}\left(k_{y}-\gamma\frac{d_{1}}{l_{B}^{2}}\right)^{2}$,
$\epsilon_{0}=\epsilon-v_{1}$ and $
Q_{\gamma}(x)=\sqrt{\frac{2}{\varrho}}e^{i\pi/4}\left(\gamma
\varrho x+\epsilon_{0}\right) $, $c_{1}$ and $c_{2}$ are
constants. It gives the other component
\begin{eqnarray}\lb{hii2}
\chi_{\gamma}^{-}&=&-c_{2}\frac{1}{k_{y}-\gamma\frac{d_{1}}{l_{B}^{2}}}\left[
2(\epsilon_{0}+\gamma \varrho x)
 D_{-\nu_\gamma}\left(-Q^{*}_{\gamma}\right)
+
 \sqrt{2\varrho}e^{i\pi/4}D_{-\nu_\gamma+1}\left(-Q^{*}_{\gamma}\right)\right]\nonumber\\
 &&
 -\frac{c_{1}}{k_{y}-\gamma\frac{d_{1}}{l_{B}^{2}}}\sqrt{2\varrho}e^{-i\pi/4}
 D_{\nu_\gamma-1}\left(Q_{\gamma}\right).
\end{eqnarray}
The components of the spinor solution of the Dirac equation
\eqref{eqh1} in region {\sf 2} and {\sf 4} can be obtained from
\eqref{hii1} and \eqref{hii2} with
$\varphi_{\gamma}^{+}(x)=\chi_{\gamma}^{+}+i\chi_{\gamma}^{-}$ and
$\varphi_{\gamma}^{-}(x)=\chi_{\gamma}^{+}-i\chi_{\gamma}^{-}$. We
have the eigenspinor
\begin{eqnarray}
 \psi_{\sf j} &=& a_{{\sf j}-1}\left(%
\begin{array}{c}
 u^{+}_{\gamma}(x) \\
  u^{-}_{\gamma}(x) \\
\end{array}%
\right)e^{ik_{y}y}+a_{\sf j}\left(%
\begin{array}{c}
 v^{+}_{\gamma}(x) \\
 v^{-}_{\gamma}(x)\\
\end{array}%
\right)e^{ik_{y}y}
\end{eqnarray}
where ${\sf j=2, 4}$ and $\gamma=\pm 1$, the function
$u^{\pm}_{\gamma}(x)$ and $v^{\pm}_{\gamma}(x)$ are given by
\begin{eqnarray}
u^{\pm}_{\gamma}(x)&=&
 D_{\nu_{\gamma}-1}\left(Q_{\gamma}\right)\mp
 \frac{1}{k_{y}-\gamma\frac{d_{1}}{l_{B}^{2}}}\sqrt{2\varrho}e^{i\pi/4}D_{\nu_{\gamma}}\left(Q_{\gamma}\right)
\\
v^{\pm}_{\gamma}(x)&=&
 \pm\frac{1}{k_{y}-\gamma\frac{d_{1}}{l_{B}^{2}}}\sqrt{2\varrho}e^{-i\pi/4}D_{-\nu_{\gamma}+1}\left(-Q_{\gamma}^{*}\right)\nonumber\\
 &&
  \pm
 \frac{1}{k_{y}-\gamma\frac{d_{1}}{l_{B}^{2}}}\left(-2i\epsilon_{0}\pm
 \left(k_{y}-\gamma\frac{d_{1}}{l_{B}^{2}}\right)-\gamma2i \varrho x\right)D_{-\nu_{\gamma}}\left(-Q_{\gamma}^{*}\right).
\end{eqnarray}
In region {\sf 2}:
\begin{eqnarray}
 \psi_{\sf 2} &=& a_1\left(%
\begin{array}{c}
 u^{+}_{1}(x) \\
  u^{-}_{1}(x) \\
\end{array}%
\right)e^{ik_{y}y}+a_{2}\left(%
\begin{array}{c}
 v^{+}_{1}(x) \\
 v^{-}_{1}(x)\\
\end{array}%
\right)e^{ik_{y}y}
\end{eqnarray}
In region {\sf 4}:
\begin{eqnarray}
 \psi_{\sf 4} &=& a_3\left(%
\begin{array}{c}
 u^{+}_{-1}(x) \\
  u^{-}_{-1}(x) \\
\end{array}%
\right)e^{ik_{y}y}+a_{4}\left(%
\begin{array}{c}
 v^{+}_{-1}(x) \\
 v^{-}_{-1}(x)\\
\end{array}%
\right)e^{ik_{y}y}
\end{eqnarray}

In the region $|x|\leq d_{1}$:
From the nature of the system under consideration, we write the
Hamiltonian corresponding to region ${\sf 3}$ in matrix form as
\begin{equation}\label{eq 20}
H_m=v_{F}\left(%
\begin{array}{cc}
  \frac{V_{2}}{v_{F}}+\frac{G_{p}}{v_{F}} & -i\frac{\sqrt{2}}
  {l_{B}}\left(\frac{l_{B}}{\sqrt{2}}\left(\partial_{x}-i\partial_{y}+\frac{e}{c}A_{3}\right)\right)\\
 i\frac{\sqrt{2}}{l_{B}}\left(\frac{l_{B}}{\sqrt{2}}\left(-\partial_{x}-i\partial_{y}
 +\frac{e}{c}A_{3}\right)\right)  &  \frac{V_{2}}{v_{F}}-\frac{G_{p}}{v_{F}}\\
\end{array}%
\right)
\end{equation}
Note that, the energy gap $G_{p}$ behaves like a mass term in Dirac equation.
Certainly this will affect the above results and leads to
interesting consequences on the transport properties of our
system. We determine the eigenvalues and eigenspinors of the
Hamiltonian $H_m$ by considering the time independent equation for the spinor
$\psi_{3}(x, y)=(\psi_{3}^{+}, \psi_{3}^{-})^{T}$ using the fact that the
transverse momentum $p_{y}$ is conserved, we can write the wave
function
$\psi_{3}(x, y)=e^{ip_{y}y} \varphi_{3}(x)$
with $\varphi_{3}(x)= (\varphi_{3}^+, \varphi_{3}^-)^{T}$, the energy being defined by
$E=\upsilon_{F}\epsilon$ leads to
\begin{equation}\label{eq 23}
H_{m}\left(%
\begin{array}{c}
  \varphi_{3}^+ \\
  \varphi_{3}^-\\
\end{array}%
\right)=\epsilon\left(%
\begin{array}{c}
  \varphi_{3}^+\\
  \varphi_{3}^-\\
\end{array}%
\right)
\end{equation}
At this stage, it is convenient to introduce the annihilation and
creation operators. They can be defined as
\begin{eqnarray}
a=\frac{l_{B}}{\sqrt{2}}\left(\partial_{x}+k_{y}+\frac{e}{c}A_{3}\right),
\qquad
a^{\dagger}=\frac{l_{B}}{\sqrt{2}}\left(-\partial_{x}+k_{y}+\frac{e}{c}A_{3}\right)
\end{eqnarray}
which obey the canonical commutation relations $[a,
a^{\dagger}]={\mathbb I}$. Rescaling our energies $G_{p}=\upsilon_{F}\mu$
and $V_{2}=\upsilon_{F}v_{2}$, \eqref{eq 23} can be
written in terms of $a$ and $a^{\dagger}$ as
\begin{equation}
 \left(%
\begin{array}{cc}
  v_{2}+\mu & -i\frac{\sqrt{2}}{l_{B}}a \\
  +i\frac{\sqrt{2}}{l_{B}}a^{\dagger}  &  v_{2}-\mu \\
\end{array}%
\right)\left(%
\begin{array}{c}
  \varphi_{3}^+ \\
  \varphi_{3}^- \\
\end{array}%
\right)=\epsilon\left(%
\begin{array}{c}
  \varphi_{3}^+ \\
  \varphi_{3}^- \\
\end{array}%
\right)
\end{equation}
which gives
\begin{eqnarray}\label{eq 25}
  &&(v_{2}+\mu)\varphi_{3}^{+}-i\frac{\sqrt{2}}{l_{B}}a\varphi_{3}^-=\epsilon\varphi_{3}^+\\
&&\label{eq 26}
  i\frac{\sqrt{2}}{l_{B}}a^{\dagger}\varphi_{3}^+ +
  (v_{2}-\mu)\varphi_{3}^{-}=\epsilon\varphi_{3}^-.
\end{eqnarray}
Injecting \eqref{eq 26} in \eqref{eq 25}, we obtain a
differential equation of second order for
 $\varphi_{3}^{+}$
\begin{equation}
\left[(\epsilon-v_{2})^{2}-\mu^{2}\right]\varphi_{3}^{+}=\frac{2}{l_{B}^{2}}a
a^{\dagger}\varphi_{3}^{+}.
\end{equation}
It is clear that $\varphi_{3}^{+}$ is an eigenstate of the number
operator $\widehat{N}=a^{\dagger}a$ and therefore we identify
$\varphi_{3}^{+}$ with the eigenstates of the harmonic oscillator
$|n-1\rangle$, namely
\begin{equation}
 \varphi_{3}^{+} \sim \mid n-1\rangle
\end{equation}
which gives for (85)
 \begin{equation}
 \left[(\epsilon-v_{2})^{2}-\mu^{2}\right]
 \mid n-1\rangle=\frac{2}{l_{B}^{2}}n\mid n-1\rangle
\end{equation}
and the associated energy spectrum is
\begin{equation}
\epsilon-v_{2}=s_{3} \epsilon_{n}=s_{3}\frac{1}{l_{B}}\sqrt{(\mu
l_{B})^{2}+2n}
\end{equation}
where we have set $\epsilon_{n}=s_{3}(\epsilon-v_{2})$ and
$s_{3}=\mbox{sign}(\epsilon_{n}-v_{2})$ correspond to positive and
negative energy solutions. For this reason we write the eigenvalues
as
\begin{equation}
\epsilon=v_{2}+s_{3}\frac{1}{l_{B}}\sqrt{(\mu l_{B})^{2}+2n}
\end{equation}
The second eigenspinor component then can be obtained from
\begin{equation}
\varphi_{3}^{-}=\frac{i\sqrt{2}a^{\dagger}}{(\epsilon-v_{2})l_{B}+\mu
l_{B}}\mid n-1\rangle=
       \frac{i\sqrt{2n}}{(\epsilon-v_{2})l_{B}+\mu l_{B}} \mid n\rangle
\end{equation}
where $ \sqrt{2n}=\sqrt{(\epsilon_n l_{B})^{2}-(\mu
l_{B})^{2}}$. We find
\begin{equation}
\varphi_{3}^{-}=s_{3}i\sqrt{\frac{\epsilon_n l_{B}-s_{3} \mu
l_{B}}{\epsilon_n l_{B}+s_{3} \mu l_{B}}} \mid n\rangle
\end{equation}
After normalization we arrive at the following expression for the positive and negative energy
eigenstates
\begin{equation}
\varphi_{3}=\frac{1}{\sqrt{2}}\left(%
\begin{array}{c}
  \sqrt{\frac{\epsilon_n l_{B}+s_{3} \mu l_{B}}{\epsilon_n l_{B}}} \mid n-1\rangle \\
  s_{3} i\sqrt{\frac{\epsilon_n l_{B}-s_{3} \mu l_{B}}{\epsilon_n l_{B}}} \mid n\rangle \\
\end{array}%
\right)
\end{equation}
Introducing the parabolic cylinder functions
$D_{n}(x)=2^{-\frac{n}{2}}e^{-\frac{x^{2}}{4}}H_{n}\left(\frac{x}{\sqrt{2}}\right)$, we  express the solution in region {\sf 3} as
\begin{equation} \psi_{\sf
3}(x,y)=b_{1}\psi_{3}^{+}+b_{2}\psi_{3}^{-}
\end{equation}
with its two components
\begin{equation}
\psi_{3}^{\pm}(x, y)=\frac{1}{\sqrt{2}}\left(%
\begin{array}{c}
 \sqrt{\frac{\epsilon_n l_{B}+s_{3} \mu l_{B}}{\epsilon_n l_{B}}}
 D_{\left(\left(\epsilon_n l_{B}\right)^{2}-(\mu l_{B})^{2} \right)/2-1}
 \left(\pm \sqrt{2}\left(\frac{x}{l_{B}}+k_{y}l_{B}\right)\right) \\
  \pm i\frac{s_{3}\sqrt{2}}{\sqrt{\epsilon_n l_{B}\left(\epsilon_n l_{B}+s_{3} \mu l_{B}\right)}}
  D_{\left(\left(\epsilon_n l_{B}\right)^{2}-\left(\mu l_{B}\right)^{2}\right)/2}
 \left(\pm \sqrt{2}\left(\frac{x}{l_{B}}+k_{y}l_{B}\right)\right) \\
\end{array}%
\right)e^{ik_{y}y}
\end{equation}

As usual the coefficients $(a_1,a_2,a_3,a_4,b_1,b_2,r,t)$ can be
determined using the boundary conditions, continuity of the
eigenspinors at each interface.

\subsection{ Transmission and reflection: presence of magnetic field}

We will now study some
interesting features of our system which are reflected in the
corresponding transmission probability. Before doing so, let us
simplify our writing using the following shorthand notation
\begin{eqnarray}
&&\vartheta_{\tau1}^{\pm}=D_{\left[(\epsilon_{n}l_{B})^{2}-(\mu
l_{B})^{2}\right]/2-1}
 \left[\pm \sqrt{2}\left(\frac{\tau d_{1}}{l_{B}}+k_{y}l_{B}\right)\right]\\
&& \zeta_{\tau1}^{\pm}= D_{\left[(\epsilon_{n}l_{B})^{2}-(\mu
l_{B})^{2}\right]/2}
  \left[\pm \sqrt{2}\left(\frac{\tau d_{1}}{l_{B}}+k_{y}l_{B}\right)\right]\\
  && f_{1}^{\pm}=\sqrt{\frac{\epsilon_{n}\pm
\mu}{\epsilon_{n}}}, \qquad
f_{2}^{\pm}=\frac{\sqrt{2/l_{B}^{2}}}{\sqrt{\epsilon_{n}(\epsilon_{n}\pm
\mu)}}\\
&& u^{\pm}_{\gamma}(\tau d_{1})=u^{\pm}_{\gamma, \tau1},\qquad
u^{\pm}_{\gamma}(\tau d_{2})=u^{\pm}_{\gamma, \tau 2}\\
&& v^{\pm}_{\gamma}(\tau d_{1})=v^{\pm}_{\gamma, \tau 1},\qquad
v^{\pm}_{\gamma}(\tau d_{2})=v^{\pm}_{\gamma, \tau 2}
\end{eqnarray}
where $\tau=\pm$. Dirac equation requires the following set
of continuity conditions
\bqr \label{eq111}
\psi_{\sf 1}(-d_2)= \psi_{\sf
2}(-d_2),\qquad
\psi_{\sf 2}(-d_1)= \psi_{\sf 3}(-d_1), \qquad
\psi_{\sf 3}(d_1)= \psi_{\sf 4}(d_1),\qquad
\psi_{\sf 4}(d_2)= \psi_{\sf 5}(d_2) \eqr
That is, requirement of the continuity of the spinor wave functions at each
junction interface gives rise to the above set of equations. We prefer to
express these relationships in terms of $2\times 2$ transfer
matrices between {\sf j}-th and ({\sf j}+1)-th regions, $\mathcal{M}_{{\sf j},{\sf j}+1}$, we
obtain the full transfer matrix over the whole double barrier
which can be written, in an obvious notation, as follows
\begin{equation}\label{syst1}
\left(%
\begin{array}{c}
  1 \\
  r_{m} \\
\end{array}%
\right)=\prod_{{\sf j}=1}^{4}\mathcal{M}_{{\sf j},{\sf j}+1}\left(%
\begin{array}{c}
  t_{m} \\
  0 \\
\end{array}%
\right)=\mathcal{M}\left(%
\begin{array}{c}
  t_{m} \\
  0 \\
\end{array}%
\right)
\end{equation}
where the total transfer matrix $\mathcal{M}=\mathcal{M}_{12}\cdot \mathcal{M}_{2
3}\cdot \mathcal{M}_{34}\cdot\mathcal{ M}_{45}$ are transfer
matrices that couple the wave function in the ${\sf j}$-th region to the
wave function in the (${\sf j} + 1$)-th region. These are given explicitly by the forms
\begin{eqnarray}
&& \mathcal{M}=\left(%
\begin{array}{cc}
  \tilde{m}_{11} & \tilde{m}_{12} \\
  \tilde{m}_{21} & \tilde{m}_{22} \\
\end{array}%
\right)\\
&& \mathcal{M}_{12}=\left(%
\begin{array}{cc}
   e^{-\textbf{\emph{i}}p_{x1} d_{2}} &e^{\textbf{\emph{i}}p_{x1} d_{2}} \\
  z_{1}e^{-\textbf{\emph{i}}p_{x1} d_{2}} & -z^{\ast}_{1} e^{\textbf{\emph{i}}p_{x1} d_{2}} \\
\end{array}%
\right)^{-1}\left(%
\begin{array}{cc}
u_{1,-2}^{+} &  v_{1,-2}^{+}\\
 u_{1,-2}^{-} & v_{1,-2}^{-}\\
\end{array}%
\right)\\
&& \mathcal{M}_{23}=\left(%
\begin{array}{cc}
 u_{1,-1}^{+} &  v_{1,-1}^{+}\\
u_{1,-1}^{-}& v_{1,-1}^{-}\\
\end{array}%
\right)^{-1}\left(%
\begin{array}{cc}
\vartheta_{1}^{+} &\vartheta_{1}^{-} \\
\zeta_{1}^{+} &\zeta_{1}^{-} \\
\end{array}%
\right)\\
&& \mathcal{M}_{34}=\left(%
\begin{array}{cc}
 \vartheta_{-1}^{+} &\vartheta_{-1}^{-}\\
  \zeta_{-1}^{+} & \zeta_{-1}^{-} \\
\end{array}%
\right)^{-1}\left(%
\begin{array}{cc}
 u_{-1,1}^{+} &  v_{-1,1}^{+}\\
 u_{-1,1}^{-} & v_{-1,1}^{-} \\
\end{array}%
\right)\\
&& \mathcal{M}_{45}=\left(%
\begin{array}{cc}
 u_{-1,2}^{+} &  v_{-1,2}^{+}\\
 u_{-1,2}^{-} & v_{-1,2}^{-}\\
\end{array}%
\right)^{-1}\left(%
\begin{array}{cc}
  e^{\textbf{\emph{i}}p_{x5} d_{2}} & e^{-\textbf{\emph{i}}p_{x5} d_{2}} \\
  z_{5} e^{\textbf{\emph{i}}p_{x5} d_{2}}  & -z_{5}^{\ast} e^{-\textbf{\emph{i}}p_{x5} d_{2}}  \\
\end{array}%
\right).
\end{eqnarray}
These will enable us to compute the reflection and transmission amplitudes
\begin{equation}\label{eq 633}
 t_{m}=\frac{1}{\tilde{m}_{11}}, \qquad  r_{m}=\frac{\tilde{m}_{21}}{\tilde{m}_{11}}.
\end{equation}
More explicitly, we have for transmission
\beq
t_{m}=
\frac{e^{id_{2}\left(p_{x1}+p_{x5}\right)}\left(1+z_{5}^{2}\right)\left(\vartheta_{1}^{-}\zeta_{1}^{+}+\vartheta_{1}^{+}\zeta_{1}^{-}
\right)}{f_{2}^{+}\left(f_{1}^{-}\mathcal{L}_{1}+if_{2}^{-}\mathcal{L}_{2}\right)+f_{1}^{+}\left(f_{2}^{-}\mathcal{L}_{3}+if_{1}^{-}
\mathcal{L}_{4}\right)}\mathcal{D}
\eeq
where the quantities $\mathcal{D}$, $\mathcal{L}_{1}$, $\mathcal{L}_{2}$, $\mathcal{L}_{3}$ and $\mathcal{L}_{4}$ are defined by
 \begin{eqnarray}
\mathcal{D}&=&
\left(u_{-1,1}^{-}v_{-1,1}^{+}-u_{-1,1}^{+}v_{-1,1}^{-} \right)
\left(u_{1,-2}^{+}v_{1,-2}^{-}-u_{1,-2}^{-}v_{1,-2}^{+}
\right)\\
 \mathcal{L}_{1}&=&
\vartheta_{-1}^{-}\zeta_{1}^{+}\mathcal{F}\mathcal{G}
-\vartheta_{1}^{-}\zeta_{-1}^{+}\mathcal{K}\mathcal{J}\\
\mathcal{L}_{2}&=&\left(\zeta_{1}^{+}\zeta_{-1}^{-}-\zeta_{1}^{-}\zeta_{-1}^{+}\right)\mathcal{F}\mathcal{J}\\
 \mathcal{L}_{3}&=&
\vartheta_{-1}^{+}\zeta_{1}^{-}\mathcal{F}\mathcal{G}-\vartheta_{1}^{+}\zeta_{-1}^{-}\mathcal{K}\mathcal{J}\\
\mathcal{L}_{4}&=&=\left(\vartheta_{1}^{+}\vartheta_{-1}^{-}-\vartheta_{1}^{-}\vartheta_{-1}^{+}\right)\mathcal{K}\mathcal{G}
\end{eqnarray}
and
\begin{eqnarray}
\mathcal{F}&=&\left[u_{1,-1}^{+}v_{1,-2}^{-}-u_{1,-2}^{-}v_{1,-1}^{+}-
z_{1}\left(u_{1,-1}^{+}v_{1,-2}^{+}-u_{1,-2}^{+}v_{1,-1}^{+}\right)\right]\\
\mathcal{G}&=&\left[u_{-1,1}^{-}v_{-1,2}^{+}-u_{-1,2}^{+}v_{-1,1}^{-}
+z_{5}\left(u_{-1,1}^{-}v_{-1,2}^{-}-u_{-1,2}^{-}v_{-1,1}^{-}\right)\right]\\
\mathcal{K}&=&\left[u_{1,-1}^{-}v_{1,-2}^{-}-u_{1,-2}^{-}v_{1,-1}^{-}-z_{1}\left(u_{1,-1}^{-}v_{1,-2}^{+}-u_{1,-2}^{+}v_{1,-1}^{-}\right)\right]\\
\mathcal{J}&=&\left[u_{-1,1}^{+}v_{-1,2}^{+}-u_{-1,2}^{+}v_{-1,1}^{+}+z_{5}\left(u_{-1,1}^{+}v_{-1,2}^{-}-u_{-1,2}^{-}v_{-1,1}^{+}\right)\right]
\end{eqnarray}

Actually the transmission probabilities $T_m$ and reflection $R_m$ for each independent mode
can be obtained using the corresponding electric current density $J$ for our system. From our previous Hamiltonian,
we can show that incident, reflected and transmitted currents take the following forms
\begin{eqnarray}
&& J_{\sf {inc,m}}=  e\upsilon_{F}(\psi_{1}^{+})^{\dagger}\sigma
_{x}\psi_{1}^{+}\\
 && J_{\sf {ref,m}}= e\upsilon_{F} (\psi_{1}^{-})^{\dagger}\sigma _{x}\psi_{1}^{-}\\
 && J_{\sf {tra,m}}= e\upsilon_{F}\psi_{5}^{\dagger}\sigma _{x}\psi_{5}.
\end{eqnarray}
These can be used to write the reflection and transmission probabilities as
\begin{equation}
  T_{m}= \frac{p_{x5}}{p_{x1}}|t_{m}|^{2}, \qquad
  R_{m}=|r_{m}|^{2}.
\end{equation}

\subsection{Numerical results: presence of magnetic field}
The physical outcome of particle scattering through the double
triangular barrier depends on the energy of the incoming particle.
We numerically evaluate the transmission probability $T_{m}$ as a
function of structural parameters of the graphene triangular double barrier
in the presence of a perpendicular magnetic field, in terms of the energy
$\epsilon$, the $y$-component of the wave vector $k_{y}$, the magnetic
field $B$, the energy gap $\mu$ and the applied potentials $v_{1}$ and
$v_{2}$. The results are shown in Figures \ref{figm1}, \ref{figm2}
and \ref{figm3}. In addition to the expected above-barrier full
transmission for some values of $\epsilon l_{B}$ and $v_{2}l_{B}$.
We note that in Figure \ref{fig01m}, when the energy is less than
the height of the potential barrier $\epsilon
l_{B}<k_{y}l_{B}+\frac{d_{1}}{l_{B}}$, we have zero transmission.
In the second interval $k_{y}l_{B}+\frac{d_{1}}{l_{B}}\leq\epsilon
l_{B}\leq v_{1}l_{B}$ the third zone contains oscillations.
Finally the interval $\epsilon l_{B}>v_{1}l_{B}$ contains the
usual over barrier oscillations and asymptotically goes to
unity at high energy. Figure \ref{fig1m} shows the transmission
spectrum for different transverse wave vector $k_{y}l_{B}$, the energy gap
$\mu l_{B}$ is zero and $v_{2}l_{B}=0$. We see that if we increase
the wave vector $k_{y}l_{B}$ the zone of zero transmission
increases according to the condition $\epsilon
l_{B}<k_{y}l_{B}+\frac{d_{1}}{l_{B}}$. In the second interval the
transmission oscillates get very close to sharp Fabry--P\'erot type of resonances and 
zero transmission regions appear as $k_{y}l_{B}$ increases. Finally in the
interval $\epsilon l_{B}>v_{1}l_{B}$ the transmission increases to its asymptotic value. 
\begin{figure}[ht]
        \centering
        \subfloat[]{
            \centering
            \includegraphics[scale=0.5]{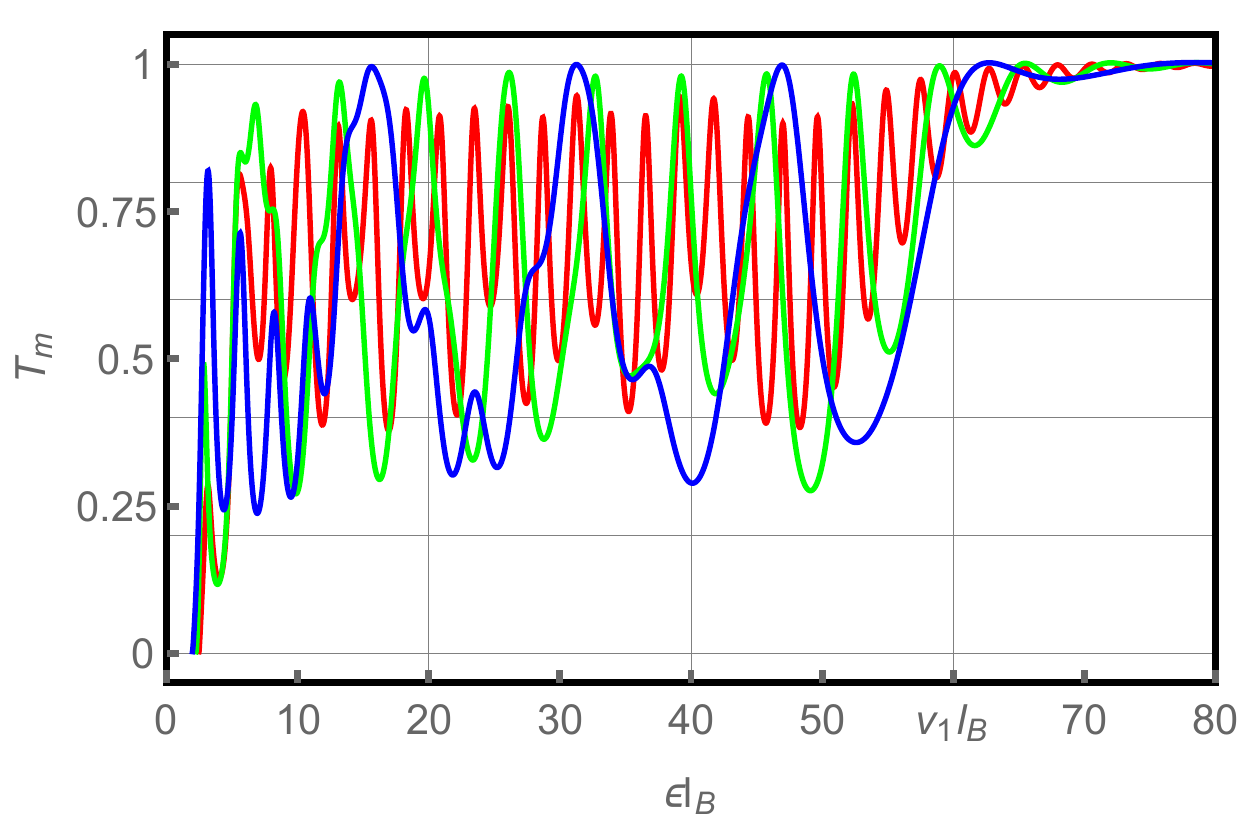}
            \lb{fig01m}\ \ \ \
        }\subfloat[]{
            \centering
            \includegraphics[scale=0.5]{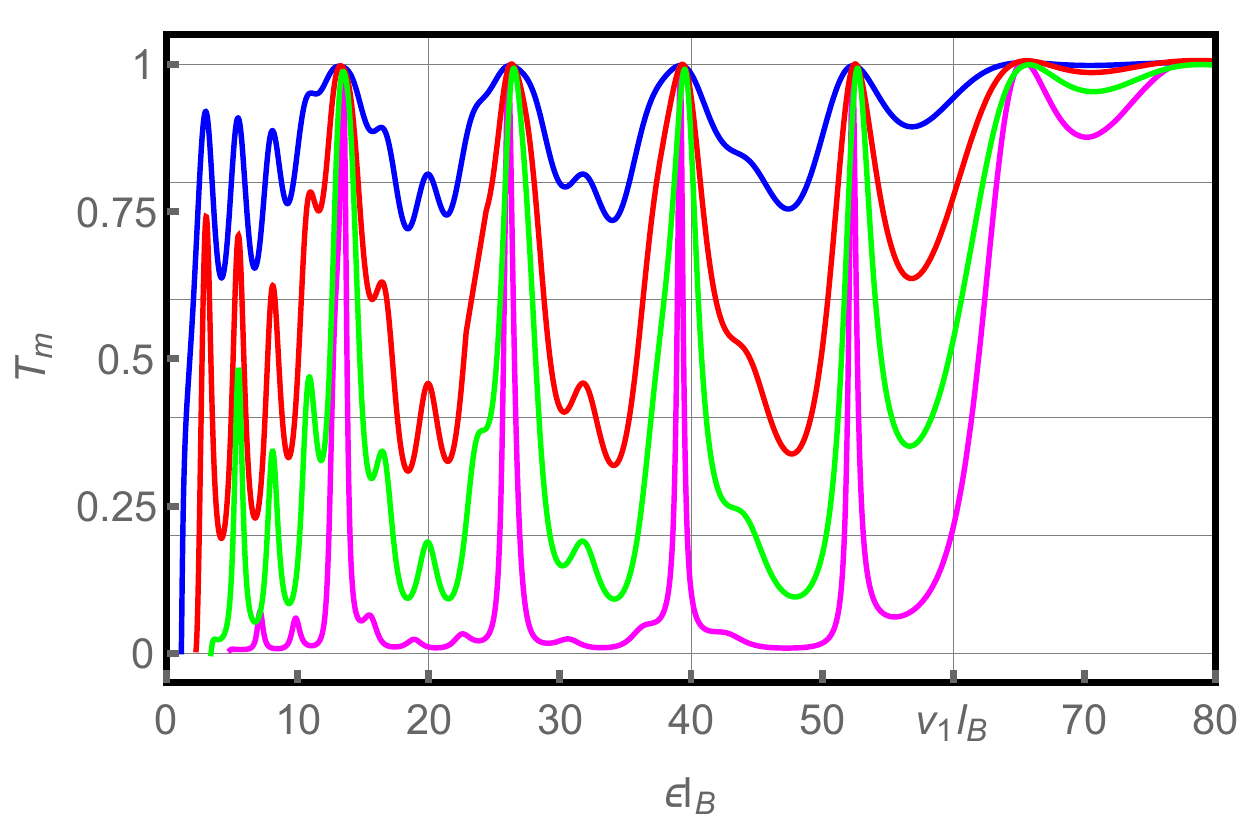}
            \lb{fig1m}}
        \caption{\sf{(color online) Transmission probability $T_{m}$ for the magnetic barrier as a function of energy
 $\epsilon l_{B}$ with $\frac{d_{2}}{l_{B}}=1.5$,
 $v_{1}l_{B}=60$, $v_{2}l_{B}=0$ and $\mu l_{B}=0$. {\color{red}{(a)}} for
 $k_{y}l_{B}=2$,
{{$\frac{d_{1}}{l_{B}}=0.12$ (blue color),
$\frac{d_{1}}{l_{B}}=0.24$ (green color) and
$\frac{d_{1}}{l_{B}}=0.6$ (red color)}}. {\color{red}{(b)}} for
$\frac{d_{1}}{l_{B}}=0.12$, {{$k_{y}l_{B}=1$ (blue
color), $k_{y}l_{B}=2$ (red color), $k_{y}l_{B}=3$ (green color)
and $k_{y}l_{B}=5$ (magenta color)}}.}}
      \lb{figm1}
    \end{figure}

On the other hand, if we keep the same well region and cancel both
the applied magnetic field and mass term in the well region, the
series of potentials behave like a simple double barrier with the
same effective mass $k_{y}$. Thus, in this case we reproduce
exactly the transmission obtained in \cite{Alhaidari}, for the
massive Dirac equation with $m = k_{y}$. Let us treat the
triangular double barrier case when $v_{2}<v_{1}$ and
$v_{2}>v_{1}$. In both cases, the transmission is plotted in
Figure \ref{figm2} and then  in Figure \ref{fig2m} $v_{2}>v_{1}$ we
distinguish five different zones characterizing the behavior of
the transmission. Indeed,
the first zone is determined by the greater effective mass,
namely $\epsilon l_{B}<k_{y}l_{B}+\frac{d_{1}}{l_{B}}$.
  The second zone identifies with the lower Klein energy zone characterized
by resonances in the interval $k_{y}l_{B}+\frac{d_{1}}{l_{B}}<\epsilon
l_{B}<v_{1} l_{B}$. Here we have full transmission at some
specific energies despite the fact that the particle energy is
less than the height of the barrier. As $d_{1}/l_{B}$ increases,
the oscillations in the Klein zone get reduced. This strong
reduction in the transmission in the Klein zone seem to suggest
the potential suppression of Klein tunneling as we increase the width of the well region,
$d_{1}/l_{B}$.
The third zone $v_{1}l_{B}<\epsilon
l_{B}<v_{2}l_{B}-k_{y}l_{B}-\frac{\mu l_{B}}{2}$ is a window where
the transmission oscillates around the value of the full
transmission.
  The fourth zone defined by $v_{2}l_{B}-k_{y}l_{B}-\frac{\mu
l_{B}}{2}<\epsilon l_{B}<v_{2}l_{B}+k_{y}l_{B}+\frac{\mu
l_{B}}{2}$ is a window where the transmission is almost zero.
 The fifth zone $\epsilon
l_{B}>v_{2}l_{B}+k_{y}l_{B}+\frac{\mu l_{B}}{2}$ contains
oscillations, the transmission converges towards unity.
Contrary to the case $v_{1}>v_{2}$, the situation where $v_{2}>v_{1}$ is shown in Figure \ref{fig3m} we distinguish fourth different zones characterizing the behavior of
the transmission. Indeed, 
compared to Figure \ref{fig2m}, the behavior in the first zone is
the same.
 Concerning the zones
$k_{y}l_{B}-\frac{d1 }{l_{B}}<\epsilon
l_{B}<v_{2}l_{B}-k_{y}l_{B}-\frac{\mu l_{B}}{2}$ and
$v_{2}l_{B}+k_{y}l_{B}+\frac{\mu l_{B}}{2}<\epsilon
l_{B}<v_{1}l_{B}$ the transmission oscillates similarly to Figure
\ref{fig2m}.
In the zone $v_{2}l_{B}-k_{y}l_{B}-\frac{\mu
l_{B}}{2}<\epsilon l_{B}<v_{2}l_{B}+k_{y}l_{B}+\frac{\mu
l_{B}}{2}$, one can see that both curves start from zero
transmission and oscillate while the valley gets wider as
$d_{1}/l_{B}$ decreases.
 Finally zone $\epsilon
l_{B}>v_{1}l_{B}$ the transmission oscillate to reach full
transmission asymptotically.

\begin{figure}[ht]
	\centering
	\subfloat[]{
		\centering
		\includegraphics[scale=0.5]{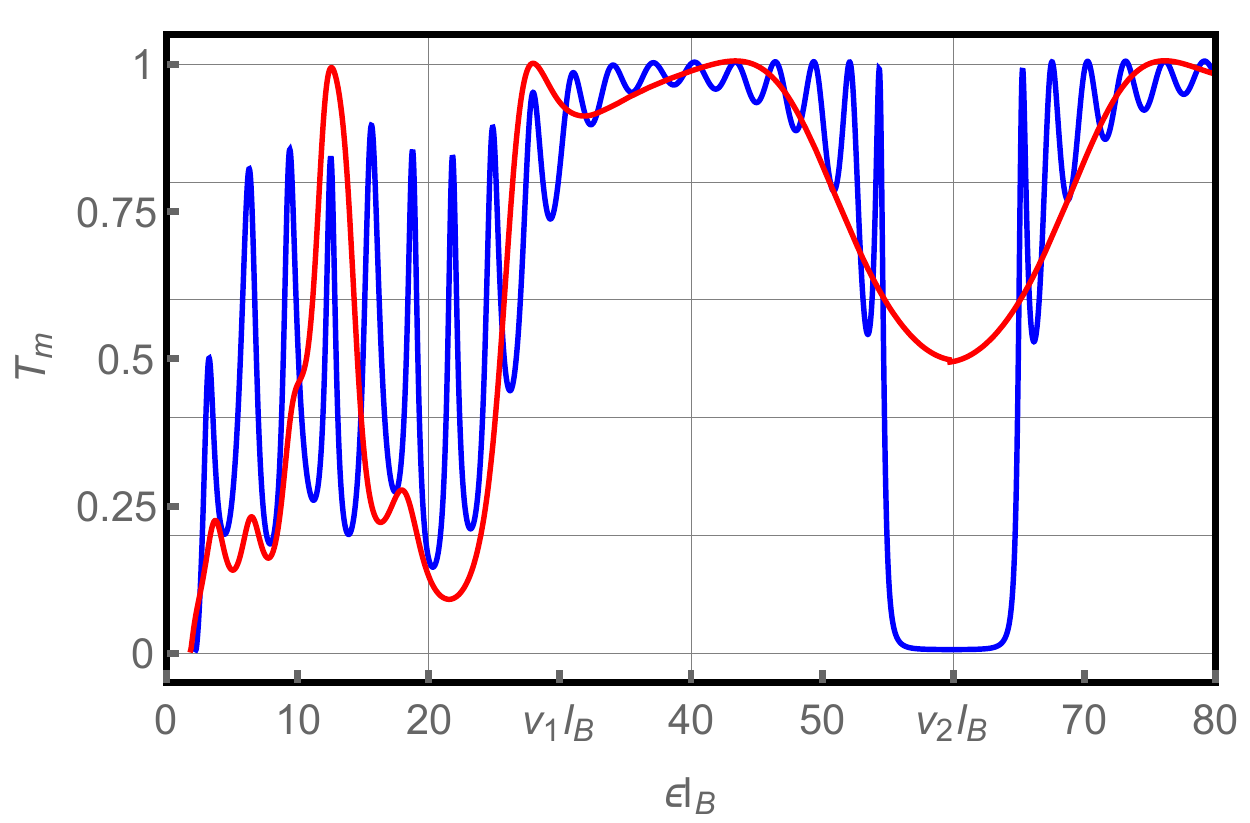}
		\lb{fig2m}
	}\subfloat[]{
		\centering
		\includegraphics[scale=0.5]{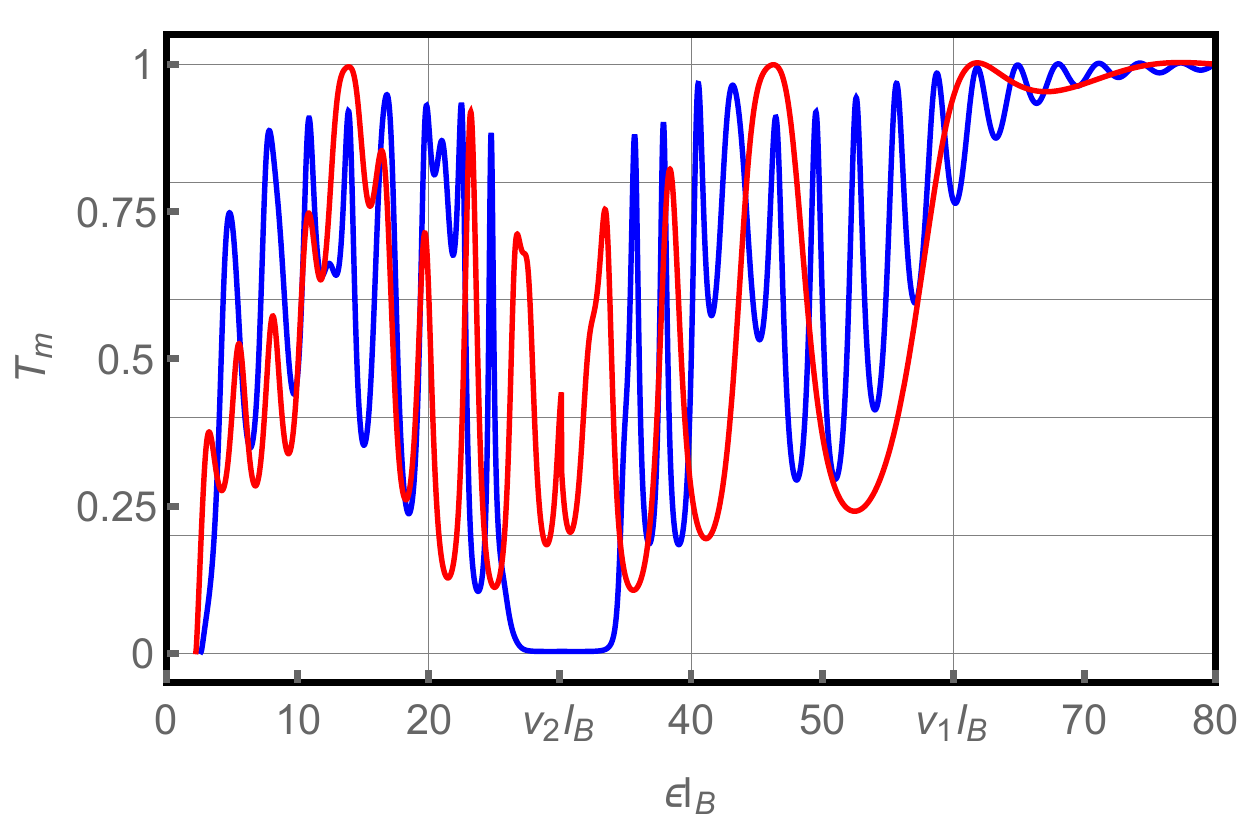}
		\lb{fig3m}}
	\caption{\sf{(color online) Transmission probability $T_{m}$ for the magnetic barrier  as a function of energy
			$\epsilon l_{B}$ with  $\frac{d_{1}}{l_{B}}=0.1$ (red color), $\frac{d_{1}}{l_{B}}=0.5$ (blue color), $\frac{d_{2}}{l_{B}}=1.5$, $\mu l_{B}=4$ and $k_{y} l_{B}=2$.
			{\color{red}{(a)}}:
			$v_{1} l_{B}=30$ , $v_{2} l_{B}=60$. {\color{red}{(b)}}: $v_{1}l_{B}=60$ , $v_{2}l_{B}=30$.}}
	\lb{figm2}
\end{figure}

\begin{figure}[h!]
\centering
\includegraphics[scale=0.5]{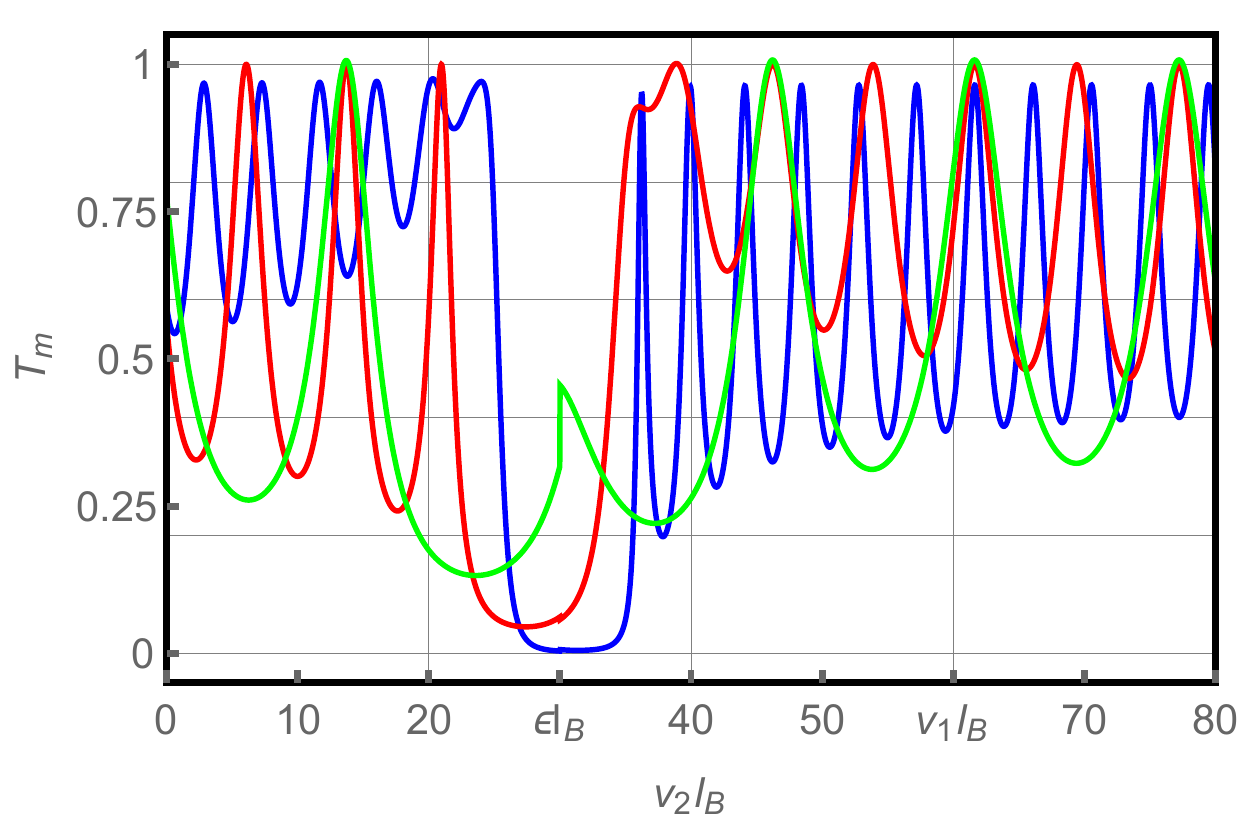}
 \caption{\sf{(Color online) Transmission probability $T_{m}$ for the magnetic barrier as a function of the potential
 $v_{2}l_{B}$ with $\frac{d_{1}}{l_{B}}=0.1$ (green color), $\frac{d_{1}}{l_{B}}=0.2$ (red color), $\frac{d_{1}}{l_{B}}=0.34$ (blue color),
 $\frac{d_{2}}{l_{B}}=1.5$, $\mu l_{B}=4$, $k_{y} l_{B}=2$,
 $v_{1}l_{B}=60$ and $\epsilon l_{B}=30$.
}}\lb{figm3}
\end{figure}

It is worth analyzing the transmission versus
the potential $v_{2}l_{B}$. In doing so, we choose a fixed value of
$d_{1}/l_{B}$ as shown in Figure \ref{figm3}. It is clear that as we increase the well width
$d_{1}/l_{B}$, hence for a fixed $v_{1}l_{B}$ and $d_{2}/l_{B}$ we will be increasing the strength of the electrostatic field, the number of oscillations increases and a zero transmission region appears.

\section{Conclusion}

We have considered a model that describes electron transmission in monolayer graphene through a triangular electrostatic double barriers in the presence of a magnetic field localized in the well region. To investigate the transport properties of our system, we have considered separately the two effects: first we considered an electrostatic double barrier alone then we added a 
uniform magnetic field effect in the well region. In both cases, we have investigated analytically and numerically the transmission probability. This has been achieved through solving the corresponding eigenvalue equation which resulted in the energy spectrum in terms of different physical parameters that control the system Hamiltonian.

Using the continuity of the wavefunctions at the interfaces
between different regions inside and outside the barriers we have
ensured conservation of the local current density and derived the
relevant transport coefficients of the present system.
Specifically, using the transfer matrix method, we have deduced
the corresponding transmission coefficient and determined how the
transmission probability is affected by various physical
parameters. In particular for the electrostatic double barrier
in the absence of magnetic field, the resonances were seen to appear in different Klein
tunneling regions. A gap region appeared for energies around the value of the elevated well potential, independently of the strength of the electrostatic field.

Subsequently, we have analyzed the same system but this time taking
into account the presence of a static magnetic field localized in the well region. Using boundary conditions, we have split the energy space into three domains and then calculated the transmission probability in each domain. In each situation, we have discussed the transmission and resonances that characterize each region and stressed the importance of our results.

Our results indicate that the Klein paradox, full transmission close to normal incidence, persists for all values the electrostatic field or equivalently the double barrier-tilting angle. From the application point of view, our current investigation suggests that tunable filtering of Dirac electrons by an electrostatic double barrier is possible for near normal incidence at all values of the electrostatic field or tilting angle. This observation could be very beneficial in designing graphene based electronic lenses.

\section*{Acknowledgments}

The generous support provided by the Saudi Center for Theoretical Physics (SCTP)
is highly appreciated. AJ and HB  
acknowledge the support of King Fahd University of Petroleum and minerals.

\end{document}